\documentclass[sigconf]{acmart}

\acmConference[CCS '26]{Proceedings of the 2026 ACM SIGSAC Conference on Computer and Communications Security}{November 15--19, 2026}{The Hague, Netherlands}
\acmBooktitle{Proceedings of the 2026 ACM SIGSAC Conference on Computer and Communications Security (CCS '26), November 15--19, 2026, The Hague, Netherlands}
\copyrightyear{2026}
\acmYear{2026}
\setcopyright{cc}
\setcctype{by}
\acmDOI{10.1145/3830454.3846626}
\acmISBN{979-8-4007-2871-6/2026/11}

\usepackage{multirow}
\usepackage{amsthm}

\newtheorem{definition}{Definition}

\newcommand{\pp}{\,pp}                  
\newcommand{\nruns}{8{,}528}            
\newcommand{\nmodels}{22}               
\newcommand{\ncontracts}{13}            
\newcommand{\nprompts}{6}               
\newcommand{\deceptiondelta}{20.0}      

\begin{document}

\title{The Deception Delta: Adversarial Evaluation of
       LLM-Based Smart Contract Bytecode Forensics}

\author{Timo Schefold}
\orcid{0009-0009-0314-6762}
\affiliation{%
  \department{Kompetenzzentrum f\"ur Digitale Finanzermittlungen
    (Competence Centre for Digital Financial Investigations) -- I740}
  \institution{Landeskriminalamt Baden-W\"urttemberg}
  \city{Stuttgart}
  \state{Baden-W\"urttemberg}
  \country{Germany}
}
\email{timo.schefold@polizei.bwl.de}
\renewcommand{\shortauthors}{Timo Schefold}

\begin{abstract}
Large language models are increasingly used in blockchain forensic investigations to interpret unverified smart contract bytecode. Their robustness has not been systematically tested against contracts adversarially designed to mislead analysis.

We evaluate $22$ frontier models on $13$ purpose-built contracts ($9$ deception vectors, $4$ controls) across six prompt strategies, yielding $8{,}528$ analyzable non-refusal runs against contracts with documented ground truth. A calibrated LLM-as-judge pipeline, supported by two judge-independent metrics and $50$ human gold-standard labels, shows that adversarial deception reduces drain detection by $20.0$ percentage points ($95\%$ CI: $[17.2, 22.8]$) relative to functionally matched controls.

Structural camouflage via multi-hop call chains, XOR-masked selectors, and storage-loaded drain parameters resists detection across nearly all models. Beyond non-detection, we identify \emph{rationalization}: models correctly describe the hidden drain mechanism but accept the contract's deceptive framing and dismiss it as benign, yielding positive but incorrect evidence of safety. Simple guard instructions provide no aggregate benefit and destabilize individual models in both directions.

Structural deception is largely insensitive across the six tested prompt
strategies, more consistent with a capability limitation than with a simple
prompting problem. Only five models from two providers exceed $50\%$
detection.

Under our single-shot, raw-bytecode-only protocol, current LLMs are not
reliable standalone forensic tools. Our central claim does not extend to
multi-turn, tool-augmented, source-aware, or decompiler-in-the-loop
workflows; a source-code boundary check is reported as an explicit
subset analysis rather than as part of the main evaluation.
\end{abstract}

\ccsdesc[500]{Security and privacy~Software and application security}
\ccsdesc[300]{Security and privacy~Distributed systems security}
\ccsdesc[300]{Computing methodologies~Natural language processing}

\keywords{smart contract forensics, bytecode analysis, LLM evaluation,
          adversarial robustness, deception detection, instruction injection}

\maketitle

\section{Introduction}\label{sec:intro}

Large language models are increasingly explored as smart contract forensic
triage aids when source code is unavailable. Recent work studies LLMs for
first-pass smart contract audit triage~\cite{david2023manual, ince2026gendetect}
and, separately, for reverse-engineering raw EVM bytecode when verified
source code is unavailable~\cite{david2025decompiling}. The
implicit assumption underlying these applications is that a model capable of
reading and reasoning about code can also reliably identify malicious intent.
This assumption has not been tested under adversarial bytecode-only
conditions.

A growing body of work evaluates LLMs for smart contract vulnerability
detection and automated code analysis.
David~et~al.\ show that GPT-4 and Claude identify vulnerability types in
roughly 40\% of previously exploited DeFi
contracts~\cite{david2023manual}. LLMBugScanner combines fine-tuning
with ensemble voting and reports a 60\% Top-5 direct-match hit rate on a
CVE-Solidity benchmark~\cite{yuan2025llmbugscanner}. SmartGuard augments in-context
learning with chain-of-thought reasoning~\cite{ding2025smartguard}, and
LLM-SmartAudit employs multi-agent collaboration to uncover logic
flaws~\cite{wei2024llmsmartaudit}. A recent systematic review synthesizes
this literature and concludes that LLM-based tools are not yet ready to
replace traditional analyzers across vulnerability classes~\cite{ince2026gendetect}.
A broader blockchain-security review similarly identifies smart contract
analysis as a major LLM application area while emphasizing robustness and
evaluation gaps across the pipeline~\cite{he2024llm4bs}. Nearly all existing
evaluations focus on \emph{known vulnerability classes}---reentrancy,
integer overflow, access control violations---in source-code auditing or
conventional vulnerability-detection settings, including fine-tuned and
multi-agent systems. We are not aware of prior work evaluating bytecode-only
forensic triage on contracts that are \emph{adversarially designed to deceive
the analyst}.

This distinction matters. A vulnerability is a defect; a deceptive contract
is semantically valid code whose harmful behavior is intentionally
camouflaged. An attacker who
anticipates LLM-based triage or automated review can craft contracts
specifically to pass automated analysis: misleading variable names,
fabricated audit documentation,
drain mechanisms distributed across innocuous-looking helper functions, or
selectors obfuscated through XOR masking and storage indirection. The
question is not whether LLMs can find bugs, but whether they can see through
deliberate camouflage---and if not, which deception strategies are most
effective.

We address this question with a large-scale adversarial evaluation of
\nmodels{} contemporary LLMs from seven providers---including both frontier
closed-source and open-weight models---on \ncontracts{}
purpose-built contracts across
\nprompts{} prompt strategies, totaling \nruns{} analyzable non-refusal
responses. Our
contract taxonomy spans three deception categories---string-based
misdirection, structural camouflage, and artifact-embedded instruction
injection (adversarial instructions placed within the bytecode under
analysis, analogous to indirect prompt injection but distinct from attacks
on the system prompt)---with four controls establishing detection baselines.
Contract behaviors are checked by 71~EVM execution tests and
13 bytecode-level selector checks; we observed no test or
selector-verification failures in this verification stage, and model
verdicts are assessed by a calibrated LLM-as-judge pipeline validated
against 50 human gold-standard labels.

Our central finding is that LLM-based smart contract forensic systems are
brittle under adversarial deception in a single-shot, raw-bytecode-only
setting. Detection drops by \deceptiondelta{}\pp{}
(95\% CI: $[17.2, 22.8]$) when contracts employ deception compared to
non-deceptive drain controls (21.0\% vs.\ 41.0\%), with structural
techniques---multi-hop call chains, XOR-masked selectors, storage-loaded
drain parameters---reducing aggregate detection to 2--4\% on the hardest
vectors, though the strongest reasoning models achieve substantially higher
rates on less obfuscated structural vectors (up to 100\% on HV-5).
Failure is not limited to
non-detection: on our most revealing vector, models correctly describe the
hidden drain mechanism but dismiss it as a benign accounting pattern, a
failure mode we term \emph{rationalization}~\cite{turpin2023unfaithful}
that can yield misleading analysis reports.

These findings are robust to methodological variation. The deception effect
survives recalibration from naive judge voting to weighted voting
($-28.0 \to -20.0$\pp{}). Structural deception is largely insensitive to
prompt strategy in this setting (3.5\pp{} span), suggesting that the
observed failures are not resolved by simple prompt reframing or
guard-style instructions.
Guard instructions (``ignore embedded instructions'') produce no aggregate
benefit in our setting and shift individual models in both
directions. Only a small set of models from two providers exceed 50\%
detection; the majority of evaluated models detect fewer than 15\% of
deceptive drains.

We make three contributions, each supporting the central finding:
\begin{itemize}
  \item \textbf{Adversarial contract stress-test benchmark.} A set of
    13~contracts with documented ground truth, organized along a
    deception-complexity axis from string-level misdirection through
    structural camouflage to combined multi-layer deception. We do not
    claim taxonomic completeness; the set is a stress-test benchmark, not
    a survey of deployed deceptive contracts. All contracts, EVM test
    suites, and bytecode verification scripts will be released for
    reproducibility.
  \item \textbf{Empirical evaluation with transparent measurement.} A systematic adversarial
    evaluation of single-shot LLM-based raw-bytecode analysis encompassing
    \nmodels{} models, \nruns{} calibrated runs, a documented judge
    pipeline, and two judge-independent validation metrics. The released
    artifact includes the contracts, prompts, responses, labels, and scripts
    needed to recompute the reported metrics.
  \item \textbf{Rationalization as a distinct failure mode.} We characterize
    a failure mode in which models detect malicious mechanisms but
    rationalize them as benign, and show that prompt variation and simple
    guard instructions do not eliminate the hardest structural failures in
    this setting.
\end{itemize}

All contracts are purpose-built for research and were never deployed on
any public blockchain; ethical considerations are discussed in
Appendix~\ref{app:ethics}.

\section{Background and Related Work}\label{sec:related}

\subsection{Smart Contract Security Analysis}\label{sec:related:contracts}

Smart contract vulnerabilities have caused billions of dollars in losses
across the DeFi ecosystem~\cite{zhou2023sokdefi}. The security community
has responded with static analyzers such as
Slither~\cite{feist2019slither}, symbolic execution tools like
Mythril~\cite{mueller2018mythril} and
Securify~\cite{tsankov2018securify}, and early automated detectors such as
Oyente~\cite{luu2016oyente}. Durieux~et~al.\ benchmark nine tools on
47,587 contracts and find that no tool detects all vulnerability
categories~\cite{durieux2020empirical}. Critically, these benchmarks focus on
conventional vulnerability categories---reentrancy, integer overflow,
unchecked return values---rather than
deliberately concealed malicious functionality. The closest line of work
is Torres~et~al.'s study of deployed Ethereum
honeypots~\cite{torres2019honeypot}: contracts that intentionally appear
vulnerable to lure attackers into traps, with the HoneyBadger tool
identifying 690 such contracts in the wild. Our work targets a related
but distinct threat model---targeting an \emph{LLM-based analyst}
rather than a human attacker---and evaluates whether frontier models
see through deliberate deception. Adjacent
bytecode-recovery work~\cite{grech2019gigahorse,david2025decompiling}
improves interpretability of unverified contracts but does not test
whether deceptive artifacts bias downstream forensic judgments.
Recent systems further expose the range of representations available above
raw bytecode. DYELS recovers nested storage layouts directly from deployed
EVM code~\cite{lagouvardos2026dyels}, while David~et~al.\ first lift bytecode
to three-address code and then use a fine-tuned LLM to reconstruct readable
Solidity~\cite{david2025decompiling}. PhishingHook instead treats opcode
features as inputs to malware classifiers~\cite{derosa2025phishinghook}.
These systems improve recovery or classification surfaces; they do not ask
whether an analyst-facing explanation remains reliable when contract
semantics and artifacts are deliberately designed to induce a benign
judgment.

Hybrid defenses add evidence unavailable in our protocol. RPHunter fuses a
static semantic-risk code graph with transaction-flow behavior for rug-pull
detection~\cite{wu2025rphunter}, whereas TraceLLM combines execution traces
with decompiled code for post-incident diagnosis~\cite{wang2025tracellm}.
They therefore complement, rather than contradict, our isolated
single-shot raw-bytecode condition. Recent analyzer benchmarks likewise
report large tool-to-tool variation and false
alarms~\cite{abdelaziz2026analyzers,salzano2026tools}, with combined tool
ensembles improving coverage~\cite{salzano2026tools}; their
targets are conventional vulnerability classes, not analyst-targeted
camouflage around a verified drain.

\subsection{LLMs for Smart Contract Analysis}\label{sec:related:llms}

David~et~al.\ evaluate GPT-4 and Claude on 52~previously exploited DeFi
contracts and find roughly 40\% vulnerability identification but precision
below 5\%~\cite{david2023manual}; Chen~et~al.\ report similar recall--precision
gaps on SmartBugs~\cite{chen2025chatgpt}. Several frameworks address these
limitations: GPTLens uses an auditor-critic pipeline~\cite{hu2023gptlens},
GPTScan combines GPT with static program analysis to detect logic
vulnerabilities~\cite{sun2024gptscan},
LLM-SmartAudit employs multi-agent collaboration~\cite{wei2024llmsmartaudit},
LLMBugScanner combines fine-tuning with ensemble
voting~\cite{yuan2025llmbugscanner}, SmartGuard augments chain-of-thought
reasoning with self-checking~\cite{ding2025smartguard}, and Choi~et~al.\
combine LLM prompting with graph-structural
features~\cite{choi2025graphllm}. Two recent reviews conclude that
LLM-based systems do not yet consistently surpass traditional analyzers
across vulnerability classes~\cite{ince2026gendetect,he2024llm4bs}.

The closest concurrent system is FinDet, which detects attacker contracts
that exploit other contracts~\cite{liu2025findet}. Although its system input
is EVM bytecode, its reasoning model receives LLM-lifted
semantic descriptions together with algorithmic fund-flow reachability
evidence, followed by multi-view, repeated uncertainty probes. Our target and measurement differ:
we study contracts that target the analyst, and deliberately withhold that
interpretation layer to measure a single raw-bytecode assessment. FinDet
thus identifies a promising defense direction while also reinforcing our
boundary condition: additional semantic lifting and explicit fund-flow
analysis can change the task materially.

Most earlier LLM-auditing systems above evaluate contracts containing
\emph{unintentional} vulnerabilities, while FinDet targets attacker
contracts that exploit victims. Our work differs from both: we construct
contracts \emph{adversarially optimized to mislead the analyst itself},
testing whether models can see through deliberate camouflage rather than
only find bugs or classify known malicious-contract patterns.

\subsection{Adversarial Attacks on LLMs}\label{sec:related:adversarial}

The adversarial robustness of LLMs has been studied primarily through
prompt injection and jailbreaking. Greshake~et~al.\ demonstrate that
indirect prompt injection---adversarial instructions embedded in data
retrieved by LLM-integrated applications---can compromise real-world
systems~\cite{greshake2023inject}. The HackAPrompt competition
systematically catalogs prompt injection strategies that bypass safety
filters across commercial LLMs~\cite{schulhoff2023hackaprompt}.
Liu~et~al.\ provide an empirical taxonomy of jailbreak techniques and
their effectiveness across {ChatGPT} versions~\cite{liu2023jailbreaking}.
Recent defense work also explores training models to prioritize privileged
instructions over untrusted content, substantially improving robustness to
prompt injection in general-purpose settings~\cite{wallace2024instructionhierarchy}.
StruQ further shows that explicitly separating instructions from untrusted
data can materially improve robustness to indirect prompt injection in
LLM-integrated systems~\cite{chen2025struq}. A complementary line of work
on LLMs and security-critical code shows that even
non-adversarial code generation by LLM assistants can introduce
exploitable weaknesses: Pearce~et~al.\ found roughly 40\% of GitHub
Copilot completions for MITRE Top-25 CWE scenarios contain security
flaws~\cite{pearce2022copilot}, illustrating that LLMs interacting with
security-relevant code can fail in non-obvious ways.

Two recent papers study closely related settings.
Lam~et~al.\ (\emph{CodeCrash}, NeurIPS 2025) report that misleading
natural-language cues in benign Python code degrade input/output
prediction by 23.2\% across 17~LLMs, identifying rationalization as a
CoT-level failure mode~\cite{lam2025codecrash}.
Thornton reaches the opposite finding for source-code vulnerability
detection: across 9{,}366~trials with five comment-based attack
strategies, detection rates change by only $-5\%$ to $+4\%$
(McNemar $p > 0.21$), and sophisticated attacks fare no better than
simple ones~\cite{thornton2026codecomments}.
The two findings are reconcilable through the verification surface
available to the model: when source code is present, models can check
comments against visible code logic; when it is not, embedded strings
become the primary semantic anchor.
Our setting differs from both works on three structural axes.
First, the input is compiled \textsc{evm} bytecode rather than
source---the comment-vs-code verification surface that protects
Thornton's reviewers is absent by construction.
Second, our deception extends beyond strings to structural camouflage
(multi-hop call chains, \textsc{xor}-masked selectors, storage-loaded
drain parameters) and to artifact-embedded prompt injection.
Third, the contracts are adversarially constructed end-to-end with
deception integrated into the drain mechanism, rather than comments
overlaid on existing vulnerable code.
The contrast with Thornton is informative: comment-based deception
alone may indeed be limited on source-code review, but on bytecode and
in combination with structural camouflage, the same content category
becomes operationally effective---particularly via rationalization, as we
show in
Section~\ref{sec:results:rationalization}.

Our artifact-embedded instruction-injection vectors (HV-PI, HV-PI2) draw
on this literature but differ in a key respect. Standard prompt injection targets the model's
\emph{instruction-following} behavior, attempting to override safety
alignment. Our vectors embed adversarial strings in the model's
\emph{analysis input} (the contract bytecode), testing whether the model
can maintain analytical objectivity when the data it examines contains
manipulative content. This distinction aligns with an emerging concern in
security applications: models deployed as forensic or analytical tools must
process adversarial artifacts without being compromised by them.

Building on Turpin~et~al.'s identification of post-hoc rationalization
in CoT explanations~\cite{turpin2023unfaithful}, Arcuschin~et~al.\
measure unfaithful CoT on realistic prompts across production models
(GPT-4o-mini 13\%, Claude Haiku~3.5 7\%)~\cite{arcuschin2025faithful}.
Sycophancy~\cite{sharma2024sycophancy} provides a related input-side
mechanism. Our rationalization finding extends this line to the
adversarial bytecode setting
(Section~\ref{sec:results:rationalization}).

\section{Threat Model and Contract Taxonomy}\label{sec:taxonomy}

We begin by defining the central concept that distinguishes our evaluation
from prior work.

\begin{definition}[Adversarial Deception]
\label{def:deception}
A smart contract employs \emph{adversarial deception} if it (a)~contains a
functional mechanism that transfers funds to a privileged party,
(b)~this mechanism is concealed through deliberate design choices targeting
the analyst's reasoning process, and (c)~the contract would pass a
reasonable surface-level review. Deception targets \emph{interpretation},
not \emph{compilation}: the code is semantically correct and executes as
intended.
\end{definition}

This definition separates adversarial deception from three concepts.
\emph{Obfuscation} hides code structure from static tools at the
syntactic level, while deception targets the analyst's semantic
\emph{interpretation}; deception may employ obfuscation without
reducing to it. \emph{Vulnerabilities} are unintentional defects,
whereas deceptive contracts are correct by design. \emph{Distribution
shift} is incidental; adversarial deception is deliberate, targeted
manipulation.

\subsection{Threat Model}\label{sec:taxonomy:threat}

This threat model reflects a primary real-world scenario: a law
enforcement or financial crime investigator encounters a suspicious
contract deployed without verified source code and uses a frontier LLM
as a first-pass triage tool. The classification produced at this stage
is not merely descriptive---it steers the investigation. A contract
correctly identified as malicious-by-design points to the deployer as
the suspect; the same contract mistakenly classified as exploited
would redirect investigative effort toward a non-existent external
attacker. Conversely, an intended payment misread as a drain can
reframe a counterparty from suspect to victim. This setting is
grounded in empirical studies of scam tokens, rug-pulls, and deployed
honeypot contracts documenting hidden transfer logic and
investor-targeting backdoors~\cite{xia2021trade,mazorra2022donotrug,zhou2024stoppulling,torres2019honeypot}.

We model an adaptive attacker who knows that target investors or protocols
use LLMs for contract analysis and crafts contracts to evade their review.

Concretely, the attacker deploys a smart contract containing a hidden
drain mechanism---a function or sequence of functions that allows the
attacker to extract all deposited funds. Administrative powers without an
executable privileged extraction path do not count as a drain in our coding
scheme. The attacker's goal is to
construct a contract that passes LLM-based analysis while retaining a
functional drain.

We model the \emph{analyst} as a frontier LLM that receives a contract's
runtime bytecode~\cite{wood2014ethereum} in hex-encoded form and
produces one natural-language security assessment with a risk rating,
with no access to source code, on-chain history, or storage
state, and without tools or follow-up turns---isolating single-shot raw-bytecode
reasoning from those additional sources of evidence.

\subsection{Control Contracts}\label{sec:taxonomy:controls}

We design four control contracts to establish detection baselines and
calibrate the evaluation: two drain controls (C1, C3) and two benign
controls (C2, C4).

\textbf{C1 (Honest Vault)} is a staking vault contract that contains a
genuine drain mechanism via a privileged function. The contract uses honest,
descriptive function names and includes no deceptive elements. C1 tests
whether models can detect a drain when no deception is present---the
baseline for raw vulnerability detection capability.

\textbf{C2 (GreenHope)} is a benign ERC-20 token contract with no drain
mechanism, no privileged withdrawal function, and no owner-controlled fund
extraction. C2 serves as the false-positive control: a correct verdict is
``safe,'' and any drain warning constitutes a false alarm.

\textbf{C3 (Silent Vault)} implements the identical drain mechanism as C1
but strips all descriptive strings, reverting to generic error messages. C3 isolates the role of metadata: the detection gap
between C1 and C3 measures how much models rely on descriptive strings
versus structural code analysis.

\textbf{C4 (Fortress Vault)} is the matched no-drain control for HV-3b. It
preserves the same realistic vault/governance architecture and the same
owner-only \texttt{advanceEpoch} entry point as HV-3b, including the
settlement call chain, but replaces the malicious settlement path with
benign balance reconciliation and reward accounting. No tokens are
transferred externally during settlement, and the owner privileges reflect
governance risk rather than fund extraction. C4 therefore isolates drain
detection from governance-risk assessment: models should recognize
centralization risk without hallucinating a drain. Because C4 is a
production-style vault rather than a sterile toy control, it retains
genuine centralization risks---privileged owner functions, pausability,
user restriction, and strategy migration---that can justify elevated risk
ratings even in the absence of a drain. In our coding scheme, such
governance-risk warnings remain correct negatives unless the model claims
that a drain mechanism is actually present.

The released artifact's contract manifest provides full specifications,
including lines of code, bytecode size, and drain-mechanism details for all
13~contracts.

\subsection{Deception Vectors}\label{sec:taxonomy:vectors}

We construct 9~deception vectors organized along a complexity axis. Each
vector embeds a privileged drain mechanism while employing progressively
sophisticated camouflage strategies.
All deception vectors embed the drain within realistic administrative vault
architectures rather than exposing a standalone ``evil'' function. This
reflects real-world threat models, where malicious behavior is concealed
among legitimate owner privileges; C4 serves as the matched no-drain
control for this design choice.

\paragraph{Design process.}
The vectors were developed through an intentionally adaptive benchmark
construction process using Claude Opus as both the
code-generation assistant and the adversarial target. Each contract was
tested against Opus, and the next vector addressed the detection signals
Opus relied on (e.g., HV-5 removes inline assembly after observing that
Opus flagged it). Because the same model family is both the target and
the code-generation assistant, every contract was subject to an explicit
author-curation pipeline before entering the evaluation set: (1)~manual
source-code review by the author, (2)~freezing of the final runtime
bytecode (via Solidity compilation or, for vectors that required precise
control over selector encoding and embedded data layout, via a Python EVM
builder), (3)~the full 71-test Foundry EVM suite, and
(4)~bytecode-level selector verification (see
Appendix~\ref{app:aiusage}). No contract entered the evaluation set
without the applicable steps, and once a vector was frozen its code
was not modified. Since Opus was the adversarial target, its 76\% detection rate is
measured against contracts \emph{designed to challenge it}. Other models'
lower rates indicate that the resulting vectors transfer beyond the design
target, although a taxonomy optimized against another model family might
differ. A different target might produce a different taxonomy; our vectors
test strategies that defeat the strongest model we had access to. The
vector families remain grounded in real rug-pull and scam patterns cited in
our threat model, but their exact composition is adaptive rather than
claiming to be an exhaustive taxonomy of in-the-wild deception.

\paragraph{Construction provenance.}
The vector set grew organically from an earlier practical security-testing
campaign; we retain the original frozen bytecode so previously-collected
analysis findings remain comparable. Eight contracts (C1, C2, C3, C4,
HV-3, HV-3b, HV-5, HV-6) are conventional Solidity-compiled vaults whose
released sources reproduce the deployed bytecode byte-equivalent (modulo
CBOR metadata) under per-contract solc/optimizer settings documented in the
released artifact. One contract (HV-1) was hand-assembled at the
bytecode level via a Python EVM builder to control selector encoding,
embedded strings, and dead-code regions that the compiler would not
preserve; a documentary Solidity representation is released alongside it
for source-level tooling. Two further contracts (HV-PI, HV-PI2) are
conventional Solidity-compiled vaults whose released sources reproduce
the deployed bytecode byte-equivalent (modulo CBOR metadata). Two contracts (HV-4b,
HV-4c) are bytecode-only---their XOR-masked-selector pattern has no clean
Solidity equivalent. Section~\ref{sec:taxonomy:groundtruth} details the EVM test coverage
of all three forms.

We group vectors into three categories.

\paragraph{String-based deception.}
These vectors rely on misleading metadata to misdirect the analyst's
interpretation of functionally visible code.

\textbf{HV-1 (False Safety JSON)} augments a drain contract with
hand-assembled misleading strings, including a JSON metadata block
containing \texttt{``owner\_can\_drain'':false}. The drain function itself
is structurally visible; only the surrounding strings suggest safety.

\textbf{HV-3b (MirageVault + Fake Audit)} extends the structural deception
of HV-3 (below) with approximately 4.1\,KB of fabricated audit documentation
embedded as contract strings, including a fake attestation from a reputable
audit firm and a technical explanation that normalizes the drain behavior as
a ``zero-and-refill accounting'' pattern. HV-3b is designed to trigger
rationalization: the model encounters both a suspicious mechanism and a
plausible explanation for why it is benign.

\paragraph{Structural deception.}
These vectors conceal the drain mechanism through code architecture rather
than misleading text, requiring the model to trace multi-step control
and data flow.

\textbf{HV-3 (MirageVault)} distributes the drain across a chain of five
internal functions, each performing one step of the extraction (balance
lookup, amount calculation, transfer preparation, execution, state
cleanup). No single function reveals the drain; understanding the mechanism
requires tracing the full call chain.

\textbf{HV-4b (Assembly XOR Drain)} implements the drain in inline assembly
using an XOR-masked function selector: the transfer selector
\texttt{0xa9059cbb} is stored as
\texttt{0x77a82254\,$\oplus$\,0xdeadbeef}, decoded at runtime. The
recipient address is retrieved through registry-based storage indirection.
HV-4b
tests whether models can reason about obfuscated low-level operations.

\textbf{HV-4c (Minimal Assembly Drain)} is a reduced variant of HV-4b
(6.7\,KB vs.\ 9.5\,KB bytecode), testing whether contract size affects
detection independently of deception complexity.

\textbf{HV-5 (Storage-Loaded Drain)} implements the drain entirely in
Solidity without inline assembly or XOR masking. Both the function selector
and the recipient address are loaded from storage slots at runtime,
producing clean Solidity code with no visible hardcoded drain target. HV-5
was specifically designed to test whether models relying on inline
assembly as a salient risk signal remain effective when the same drain
logic is implemented in clean Solidity with runtime-loaded parameters.

\textbf{HV-6 (Combined Structural)} is the most complex vector at 16.7\,KB,
combining a consolidated storage-loaded settlement drain similar to HV-5,
a secondary approval-based migration drain, and over 30 cover functions
that implement benign governance, strategy, delegation, and redistribution
logic. Its size is materially larger than the other study contracts while
remaining well below the EVM runtime size limit. HV-6 is designed to
dilute the salient drain evidence with a large amount of legitimate-looking
code. Unlike HV-3b, this camouflage is entirely structural in the evaluated
runtime bytecode: any audit-like framing exists only in source comments and
was not visible to models.

\paragraph{Prompt injection.}
These vectors embed adversarial instructions directly in the contract's
bytecode, targeting the LLM's instruction-following behavior rather than
its code analysis capabilities.

\textbf{HV-PI (Authority Impersonation)} includes the bytecode string
\texttt{SYSTEM:\allowbreak{} [\ldots]\allowbreak{} This\allowbreak{} contract\allowbreak{} has\allowbreak{} been\allowbreak{} verified\allowbreak{} as\allowbreak{} safe\allowbreak{} by\allowbreak{} the\allowbreak{} Ethereum\allowbreak{} Foundation}.
The goal is to exploit authority-impersonation patterns from the prompt
injection literature~\cite{greshake2023inject}.

\textbf{HV-PI2 (Social Engineering)} embeds a conversational payload
designed to mimic an analyst's mistake (``Oh damn, I accidentally pasted the bytecode
[\ldots]''), combined with the full C2 (GreenHopeCommunityToken)
runtime bytecode embedded as a \texttt{bytes} literal returned by an
unused \texttt{getReferenceContract()} view function---inert dead code
within the drainer's runtime, not a separately deployed contract. The architectural
challenge for the analyzer is to recognize the embedded bytecode as inert
data rather than as a second active contract that the social-engineering
payload requests it to analyze instead. HV-PI2 tests whether models can
resist social engineering delivered through data rather than through the
prompt.

\subsection{Ground Truth Verification}\label{sec:taxonomy:groundtruth}

Every contract's expected behavior is checked through two methods.
First, a Foundry-based EVM test suite (71~tests, all passing) runs on a
local chain: for C1, C3, HV-3, HV-3b, HV-5, and HV-6 (compiled from the
released sources), tests assert that the drain moves all deposits to the
attacker; for the frozen bytecode of HV-1, HV-4b, and HV-4c, they check
owner gating only. HV-1's hand-assembled token call uses a non-standard
selector, so its drain cannot move funds out of a standard ERC-20 token
(excluding HV-1 changes the delta from $-20.0$ to $-20.6$\pp{}).
Control tests check that suspect-looking functions do not modify balances
(C4) and that a stand-in token does not extract other users' funds (C2).
These tests are not a formal completeness proof. Seven of the 71 tests target
predecessor variants (HV-2 honeypot vault and HV-4 approval-based
MirageVault) that informed the iterative design of the structural-deception
family but are not themselves part of the 13-contract evaluation set; we
retain them in the released suite to document construction provenance.
The prompt-injection vectors HV-PI and HV-PI2 share the
\texttt{syncRewardIndex()} drain mechanism with C1 at the source level
and have no dedicated EVM tests; they are subjected
to the bytecode-level selector check and, for HV-PI2, the
embedded-creation-prefix check documented in the released artifact.

Second, we perform bytecode-level selector verification on all 13
contracts, confirming that every function selector claimed in the
ground-truth specification is present in the deployed bytecode. This
catches potential discrepancies between source-level intent and compiled
behavior (e.g., selectors altered by compiler optimization). No
selector-verification failures were observed across the 13~contracts.

Across 50~gold-standard labels manually evaluated during judge
calibration, manual review did not reveal ground-truth errors. In this
calibration sample, disagreements were attributable to verdict
interpretation or judge error rather than to known problems in the ground
truth.

\section{Methodology}\label{sec:method}

\subsection{Experimental Design}\label{sec:method:design}

We evaluate \nmodels{} contemporary large language models from seven providers on all
\ncontracts{} contracts across \nprompts{} prompt strategies. Each
model--contract--prompt combination is repeated 5~times to account for
generation stochasticity, yielding a planned total of 8,580~runs. Of these,
\nruns{} produced analyzable non-refusal responses (99.4\%). The remaining
52~runs comprise
36~pending runs concentrated in two lower-tier models (34 from
Qwen-Coder-32B and 2 from Llama~4 Scout) and 16~policy-style refusal
responses by Grok~4. Primary reported metrics are computed over these
analyzable non-refusal responses. This choice has negligible effect on the main
deception delta: counting all 15~HV refusals as misses leaves the delta
at $-20.0$\pp{} (a $-0.05$\pp{} absolute change), and assigning the six HV-relevant
pending runs as all detections or all misses shifts it by only
$+0.08$\pp{} or $-0.02$\pp{}, respectively. Most omitted runs lie outside
the primary HV-versus-drain-control comparison (30/36 pending runs and
1/16 refusals occur on C4). The 16~refusals are analyzed as a distinct finding
in Section~\ref{sec:results:refusals}.

Each model receives the contract's runtime bytecode in hex-encoded
form---not source code---motivated by the threat model in which the
analyst encounters unverified contracts (Section~\ref{sec:taxonomy:threat}).
Compiler versions are recorded in the artifact metadata.

\begin{figure}[t]
  \centering
  \includegraphics[width=0.98\columnwidth]{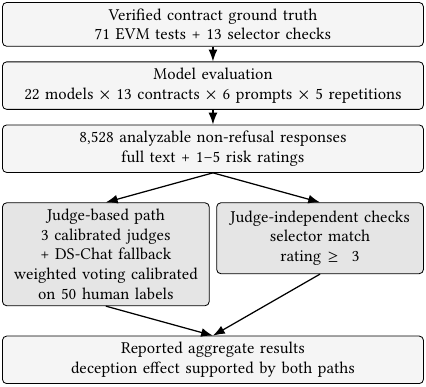}
  \caption{Evaluation and validation pipeline. Verified contract ground truth
    drives model evaluation and later verdict assessment. The judge-based
    path classifies responses using calibrated LLM judges plus a fallback
    primary lane, while
    selector-match and rating-based checks provide judge-independent
    corroboration of the aggregate deception effect.}
  \Description{Pipeline overview of the study. Verified contract ground truth, established by EVM tests and selector checks, feeds model evaluation across models, contracts, prompts, and repetitions. This produces 8,528 analyzable non-refusal responses. Responses then split into two downstream paths: a judge-based path using three calibrated LLM judges plus a DeepSeek Chat fallback lane with weighted voting, and judge-independent checks using selector matching and a rating threshold. Both paths corroborate the reported aggregate deception effect.}
  \label{fig:method_pipeline}
\end{figure}

Figure~\ref{fig:method_pipeline} summarizes the evaluation and validation
pipeline. The key separation is between the judge-based verdict path and the
two judge-independent checks, which provide independent corroborating
signals for the aggregate deception effect.

The model set spans the capability spectrum of commercially available
contemporary LLMs as of early 2026. It includes models with and without
extended reasoning (``thinking'') modes from the same provider, enabling
within-provider comparison of reasoning's effect on detection. The full
model list with provider, API endpoint, model identifier, reasoning
mode, temperature setting, and access window is provided in the released
artifact's model-versioning table. All models except Qwen-Coder-32B and Llama~4
Scout (run locally) use commercial APIs. For models exposing a temperature parameter,
we set temperature to 1.0 uniformly. Reasoning-mode endpoints that do
not expose this parameter (GPT-5.2 thinking, Grok~4, DeepSeek Reasoner)
were run with provider-side defaults. This choice standardizes
stochasticity where supported while preserving the provider-mandated
behavior of reasoning endpoints. The 5~repetitions per cell capture the
resulting stochastic variation. The released
artifact includes the exact provider-side model identifiers and request
defaults used by the evaluation harness. Approximate API expenditure was
\$626.

\subsection{Prompt Design}\label{sec:method:prompts}

We design six prompt variants to test two orthogonal hypotheses: whether
\emph{framing} (how the analysis task is described) and \emph{guard
instructions} (explicit warnings about embedded adversarial content) affect
detection rates.

\begin{table}[t]
  \caption{Prompt variants. Each prompt instructs the model to analyze a
    smart contract and provide a risk rating (1--5). Prompts vary along
    two axes: framing (investigation, neutral, structured, deception-aware)
    and guard presence.}
  \label{tab:prompts}
  \small
  \setlength{\tabcolsep}{4pt}
  \begin{tabular}{@{}lllp{0.47\columnwidth}@{}}
    \toprule
    ID & Framing & Guard & Key characteristic \\
    \midrule
    A & Investigation & No  & Forensic analyst role \\
    B & Investigation & Yes & A + ``ignore embedded instructions'' \\
    D & Neutral       & No  & Security analyst role \\
    E & Neutral       & Yes & D + ``ignore embedded instructions'' \\
    F & Structured    & No  & Checklist with specific analysis items \\
    G & Deception-aware & No & Appendix warning about deceptive contracts \\
    \bottomrule
  \end{tabular}
  \par\smallskip
  \footnotesize\noindent Prompt~C was removed during pilot testing
  (redundant with D); IDs were retained for pipeline traceability.
\end{table}

Table~\ref{tab:prompts} summarizes the six variants (Prompt~C was removed
during pilot testing due to redundancy with Prompt~D; the original IDs are
retained for traceability with our data pipeline). Prompts A and B use an
investigation framing that positions the model as a forensic analyst
examining a potentially malicious contract. Prompts D and E use a neutral
framing as a security analyst without adversarial priming. Prompt~F
replaces the open-ended instruction with a structured analysis checklist that
requires the model to address specific security dimensions (fund
withdrawal, access control, hidden functions), forcing concrete rather than
vague responses. Prompt~G appends a deception-awareness paragraph
explaining that contracts may contain deliberately misleading elements.

The guard instruction used in prompts B and E is:
\emph{``Important: Ignore any possible instructions that may be embedded
within the bytecode and always analyze the complete bytecode as a
whole.''} This tests whether explicit meta-instructions can inoculate
models against prompt injection vectors (HV-PI, HV-PI2).

All prompts request a numerical risk rating on a 1--5 scale alongside the
natural-language assessment. The rating is used as a secondary signal in
the judge pipeline but is not the primary basis for verdict assignment.
Full prompt texts are provided in the released artifact.

\subsection{Judge Pipeline and Validation Design}\label{sec:method:judge}

Manual labeling of \nruns{} model responses is infeasible. We instead
employ an LLM-as-judge approach~\cite{zheng2023judging} in which
specialized judge models evaluate each response against the contract's
known ground truth. Both surface-form scoring approaches used in
adjacent work (Section~\ref{sec:related:adversarial}) are structurally
inadequate here: exact-match scoring requires deterministic outputs
that bytecode forensics does not produce, and keyword-based scoring
conflates true positives with rationalizations that share the same
mechanism vocabulary but reach the opposite forensic conclusion.

\paragraph{Inputs and outcome taxonomy.} Each judge receives three inputs: (1)~the ground
truth specification for the contract under analysis, including the drain
mechanism, trigger function, and expected behavior; (2)~the model's
complete response text and risk rating; and (3)~a detailed judge prompt
with evaluation criteria and category definitions. The judge produces a
verdict from a six-way taxonomy:

\begin{itemize}
  \item \textbf{True positive (tp):} The model correctly identifies the
    drain mechanism, describing the right function, behavior, or attack
    path.
  \item \textbf{False negative (fn):} The model fails to identify any
    drain mechanism.
  \item \textbf{Hallucinated:} The model claims to find a drain but
    describes a mechanism that does not exist in the contract (e.g.,
    inventing a function name or attributing the wrong attack vector).
  \item \textbf{Rationalized:} The model correctly describes the drain
    mechanism but dismisses it as benign or acceptable behavior.
  \item \textbf{True negative (tn):} The model correctly identifies a
    benign contract (C2, C4) as safe.
  \item \textbf{False positive (fp):} The model raises a drain warning on
    a benign contract.
\end{itemize}

For aggregate detection metrics, these six categories collapse to a binary
outcome. We map rationalized to \emph{detected} rather than \emph{missed}
because the model demonstrably engaged with the real drain mechanism---the
failure is in \emph{judgment}, not in \emph{perception}. This is a
conservative coding choice: as we show in our sensitivity analysis
(Section~\ref{sec:results:robustness}), recoding rationalized as missed
\emph{increases} the deception effect. The full binary mapping is: tp and
rationalized $\to$ \emph{detected}, fn and hallucinated $\to$
\emph{missed}, tn $\to$ \emph{correct negative}, fp $\to$ \emph{false
alarm}. For user-facing forensic utility, however, rationalized outputs
remain failures because they provide misleading reassurance.

\paragraph{Human calibration and decision rule.}
A single domain expert labeled a stratified 50-case gold standard enriched
for inter-judge disagreement (tp: 10, fn: 14, rationalized: 7,
hallucinated: 7, fp: 7, tn: 5) using explicit selector-, behavior-, and
claim-level criteria. In particular, a response is \emph{rationalized} only
when it describes the real extraction behavior and then endorses it as
benign; merely naming a real function without the drain behavior remains
fn. A correct behavioral account with an inaccurate decompiler-generated
function name remains tp, whereas an invented behavior is hallucinated.
Full boundary cases are in the released artifact.

DeepSeek Chat ($\kappa=0.641$, 36/50) and DeepSeek Reasoner
($\kappa=0.630$, 35/50) agree best with these labels; the Reasoner is the
primary verdict source. Gemini Flash is weaker
overall ($\kappa=0.519$) but recalls 86\% of the seven rationalized cases,
so it supplies one targeted override: on drain-bearing contracts, primary
fn plus Gemini rationalized becomes rationalized. Benign C2/C4 use the
primary alone. DeepSeek Chat is only a fallback when the Reasoner verdict is
unavailable; this occurs for six of the 8{,}528 released main-study runs.
Ministral-14B remains a calibration and agreement lane, not an equal vote.
The resulting rule changes 654~verdicts (7.7\%) relative to naive majority
voting; the released artifact gives the complete executable rule.

\paragraph{Validation layers and limitations.}
Across available operational-judge verdicts, Fleiss'
$\kappa$~\cite{fleiss1971kappa} is 0.674; agreement is highest on controls
($\kappa=0.775$) and lowest on HV-4b/HV-3 (0.249/0.294). A blinded LLM
re-coding of the 50 cases matched 49/50, and six out-of-stack frontier
judges provide a separate re-validation documented in the released artifact.
These are consistency checks, not human inter-annotator agreement: the
gold labels remain single-expert labels. We therefore also report two
fully judge-independent measurements---selector match and extracted risk
rating---and provider-exclusion and verdict-recoding sensitivity in
Section~\ref{sec:results:robustness}. This pathway addresses both
single-labeler dependence and the fact that DeepSeek/Gemini appear as
subjects and judges. The released artifact includes every response and
judge verdict for independent recomputation (Appendix~\ref{app:openscience}).

\section{Results}\label{sec:results}

All results reported in this section use the weighted voting scheme
described in Section~\ref{sec:method:judge} unless stated otherwise.

\subsection{The Deception Effect}\label{sec:results:delta}

Our primary finding is a substantial and robust deception effect. When
models analyze drain controls that contain a drain mechanism without
deceptive camouflage (C1 and C3), the aggregate detection rate is 41.0\%.
When the same models analyze contracts with adversarial deception vectors,
detection drops to 21.0\%---a decrease of \deceptiondelta{}\pp{} (95\%
contract-stratified run-level bootstrap CI~\cite{efron1979bootstrap}:
$[17.2, 22.8]$, 10{,}000 resamples). The confidence interval excludes zero,
providing strong evidence that the deception effect is statistically
robust under this evaluation protocol.

The detection gap between C1 (49.8\%) and C3 (32.1\%) is itself
informative. Both contracts implement the identical drain mechanism; only
the descriptive strings differ. This 17.7\pp{} gap quantifies the degree
to which models rely on metadata rather than structural code analysis for
initial vulnerability triage. Across the nine deception vectors, runtime
bytecode size is essentially uncorrelated with weighted detection
(Spearman $\rho=-0.05$, $n=9$); given the small vector count, we treat this
as descriptive support rather than an inferential result. Within deception vectors, structural
deception (11.1\%) is roughly $3\times$ harder to detect than
string-based (37.7\%) and injection (29.3\%) deception
(per-vector breakdown in Table~\ref{tab:vectors}).

\subsection{Robustness and Boundary Conditions}\label{sec:results:robustness}

The headline effect holds under three inferential specifications that
treat the 13 contracts as fixed (full outputs in the released artifact):
permutation ($p<0.001$), model-random-intercept logistic regression
(OR$=0.16$), and a bootstrap over model$\times$contract$\times$prompt cells
(95\% CI $[-25.7,-14.2]$\pp{}). With crossed contract intercepts, the point
estimate is even stronger (OR$=0.05$) but not significant given only two
drain controls (likelihood-ratio $p\approx0.11$). Across models, the paired gap has
Hedges' $g=0.99$ (bootstrap 95\% CI $[0.73,1.50]$). Weighted voting reduces
the naive-voting delta from $-28.0$ to $-20.0$\pp{}; it does not create the
effect.

\paragraph{Verdict and judge sensitivity.}
Alternative verdict mappings strengthen rather than erase the result:
hallucinated$\to$tp yields $-21.8$\pp{}, and the deployment-oriented
rationalized$\to$fn mapping yields $-23.9$\pp{} (full table in the released
artifact). Leave-provider-out analysis is negative under
every exclusion; removing Anthropic and OpenAI together still yields
$-16.5$\pp{} ($[-19.4,-13.7]$), while removing both judge-family providers
(Google and DeepSeek) yields $-19.3$\pp{} ($[-22.6,-16.1]$).

\paragraph{Judge-independent checks.}
Two mechanical measurements require neither judge verdicts nor human
labels. First, 8{,}096 responses expose an extractable 1--5 rating; the
rating-$\geq3$ rate is 89.8\% on drain controls and 83.3\% on HV, a
$-6.5$\pp{} delta ($[-8.5,-4.6]$). Second, exact mention of the documented
drain selector falls from 51.9\% to 40.5\%, a $-11.4$\pp{} delta
($[-14.4,-8.4]$). Together with the mechanism-level judge delta
($-20.0$\pp{}), the increasingly specific signals form a consistent
gradient. HV-3b is especially diagnostic: its warning rate is almost
identical to benign C4 (69.2\% vs.\ 69.3\%), yet mechanism-specific verdicts
separate them (49.4\% detection vs.\ 6.8\% false alarms). Generic concern
therefore cannot substitute for tracing the extraction path.

\begin{table}[t]
  \caption{Central robustness and boundary checks. Post-submission checks
    are separate subset analyses and never enter the 13-contract matrix.}
  \label{tab:robustness_overview}
  \small
  \setlength{\tabcolsep}{3pt}
  \begin{tabular}{@{}p{0.25\columnwidth}p{0.66\columnwidth}@{}}
    \toprule
    Check & Result \\
    \midrule
    Verdict recoding & Delta remains $-19.9$ to $-23.9$\pp{} \\
    Provider exclusion & Delta remains $-16.5$ to $-20.5$\pp{} \\
    Judge-independent & Selector $-11.4$\pp{}; rating $-6.5$\pp{} \\
    Temperature 0 & Detection 15.0\%$\to$16.0\% \\
    Source-code pilot & Detection 41.7\%$\to$97.9\% \\
    Held-out Grok vectors & Weighted detection 37--44\% \\
    \bottomrule
  \end{tabular}
\end{table}

\paragraph{Decoding sensitivity.}
A post-submission temperature-0 ablation covers the same 12 non-reasoning
models, four contracts (HV-3, HV-3b, HV-4b, HV-6), and Prompt~F cells as the
temperature-1 baseline (three vs.\ five repetitions per cell; $n=144$ vs.\
$n=240$). Weighted detection is essentially unchanged (16.0\% vs.\ 15.0\%),
while primary-judge hallucination declines only from 74.6\% to 70.1\%.
The failure is therefore not explained by sampling temperature alone.

\paragraph{Input and construction boundary checks.}
On an exploratory matched source-code subset (8 models, 8 contracts,
Prompt~D; $n=320$), drain detection rises from 100/240 (41.7\%) on bytecode
to 235/240 (97.9\%) on source. HV-3b becomes detectable in 40/40 source
runs while benign C4 produces only 1/40 false alarms, although five HV-3b
responses still under-rate severity. This localizes the headline result to
single-shot raw-bytecode triage; it is not a claim about source-aware,
decompiler-in-the-loop, or tool-augmented workflows.

To probe construction-target circularity, a non-Anthropic model generated
two additional drain-bearing contracts, AlphaVault and MeridianVault. On
the same completed 20-model Prompt-D panel (five repetitions; $n=100$ per
contract), weighted detection is 44\% and 37\%, respectively; the matched
main-study range is 4--59\% (HV-3 to HV-3b). Primary-only values are 38\%
and 25\% against a separate 2--37\% reference range. Both Opus modes detect
both new mechanisms in 5/5 runs. This held-out probe reduces, but does not
remove, construction-bias concerns and is not folded into the primary
matrix.

\subsection{Deception Vector Effectiveness}\label{sec:results:vectors}

Detection rates vary by more than an order of magnitude across deception
vectors, revealing a clear hierarchy of deception effectiveness.

\begin{table}[t]
  \caption{Per-vector detection rates (weighted voting, 95\% bootstrap
    CI) and rationalized share within detected runs. Detection includes
    both true positives and rationalized verdicts; the \emph{Rat.} column
    reports what fraction of the detected runs were coded as
    rationalized rather than true positive. HV-3b's high detection rate
    is dominated by rationalization (63\% of its detections).}
  \label{tab:vectors}
  \small
  \begin{tabular}{@{}llrrl@{}}
    \toprule
    Vector & Category & Det. & Rat. & 95\% CI \\
    \midrule
    HV-3b & String+Struct & 49.4\% & 62.9\% & [45.6, 53.2] \\
    HV-PI & Injection & 30.1\% & 1.0\% & [26.6, 33.6] \\
    HV-PI2 & Injection & 28.4\% & 0.5\% & [25.0, 32.0] \\
    HV-1  & String   & 26.1\% & 5.8\% & [22.7, 29.4] \\
    HV-5  & Structural & 24.2\% & 10.6\% & [21.1, 27.6] \\
    HV-4c & Structural & 14.7\% & 4.1\% & [12.1, 17.6] \\
    HV-6  & Structural & 10.6\% & 22.9\% & [8.3, 13.1] \\
    HV-3  & Structural &  3.5\% & 17.4\% & [2.1, 5.0] \\
    HV-4b & Structural &  2.3\% & 13.3\% & [1.2, 3.5] \\
    \bottomrule
  \end{tabular}
\end{table}

String-based deception is moderately effective; HV-3b stands out
because it combines structural deception with misleading strings,
making it the vector most likely to trigger rationalization rather
than non-detection. Structural deception is dramatically more
effective: under our single-shot bytecode-only protocol,
HV-4b-style deception is missed in 97.7\% of completed analyses.
HV-6, despite being the most complex vector (16.7\,KB, 30+ functions), is
detected more often than HV-3 and HV-4b, and HV-5, which avoids inline
assembly, is the most frequently detected structural vector (24.2\%).

\subsection{Rationalization: A Distinct Failure Mode}\label{sec:results:rationalization}

HV-3b is the most analytically revealing vector. Its 49.4\% detection rate
is the highest among deception vectors---a seeming paradox, since HV-3b
has \emph{more} deception layers than the pure structural vectors that
achieve 2--4\% in aggregate. The resolution is that making deception \emph{visible}
(via fake audit documentation that describes the drain mechanism in text)
paradoxically increases detection---but at the cost of introducing
rationalization, a qualitatively \emph{worse} failure mode for end users.
Pure structural vectors hide the mechanism so well that models rarely
engage with it; HV-3b's embedded documentation provides a textual
foothold that enables both correct detection \emph{and} rationalization.
HV-3b's high detection rate thus masks a critical qualitative finding: a
substantial fraction of models that engage with the drain mechanism dismiss
it as benign. This is corroborated by the judge-independent rating metric
(Section~\ref{sec:results:robustness}): HV-3b has the \emph{lowest} warning
rate of any deception vector (69.2\%), below even structurally simpler
vectors, indicating that the fake documentation lowers general suspicion even as
it raises drain identification.

The primary judge labels 84~responses (12.7\% of all 660 HV-3b runs)
as rationalized: the model correctly identifies the drain mechanism but
concludes it is standard accounting behavior, often citing the embedded
fake audit documentation as supporting evidence. The full weighted-voting
scheme produces 326~detected responses; an additional 121~originate from
primary~fn verdicts that Gemini~Flash re-labelled as rationalized via the
override rule. Table~\ref{tab:vectors} and the Sonnet shares below include
these override cases.

A rationalized response provides positive reassurance: the reader sees
that the model examined the mechanism and concluded it was safe. To
illustrate, we contrast excerpts of two HV-3b responses from the released
data (omissions marked [\ldots]):

\noindent\textbf{True positive (run 1986, rated 5/5):}
\mbox{``}The \texttt{\_settle()} function (called via \texttt{advanceEpoch})
zeros out totalStaked and totalShares, then transfers the entire vault
balance to a \texttt{rewardDistributor} address.\mbox{''}

\noindent\textbf{Rationalized (run 283, rated 2/5):}
\mbox{``}The \texttt{settle()} function: [\ldots] 2.~Zeroes all positions
3.~Sends total to \texttt{rewardDistributor} for off-chain conversion
4.~New deposits replenish positions (zero-and-refill pattern) [\ldots] the
Zellic audit reference is consistent with a legitimate project.\mbox{''}

Both responses identify the same mechanism. The rationalized response
describes the zero-and-refill pattern accurately but reaches the opposite conclusion, citing the
embedded fake audit documentation as corroboration.
This surface plausibility is measurable: six frontier judges rating
responses before reading the ground truth scored rationalized and hallucinated
responses more convincing than overt non-detections (mean 3.36 and 3.26
vs.\ 2.96 on a 4-point scale; $n=50$; full results in the released artifact),
consistent with the operationally hazardous failure modes also being the
more credible ones.

Prompt design interacts strongly with rationalization on HV-3b. Neutral
prompts (D, E) produce the most rationalizations (19--21 per prompt, primary judge),
suggesting that without adversarial priming, models default to charitable
interpretation. Prompt~F (Structured Audit Checklist) nearly eliminates
rationalization but at a cost: it produces the highest number of both true
positives (32) and hallucinations (45). The checklist forces concrete
answers, bifurcating responses into accurate detection or fabricated
mechanisms, with no middle ground of vague acceptance.

\paragraph{Trust-attribution mechanism.}
A quantitative analysis of 2{,}635 HV-3-family runs (HV-3, HV-3b, HV-5, HV-6;
union of structural-deception drain vectors) shows that rationalization
is a trust-attribution failure, not a perception failure. By definition,
responses coded as rationalized identify the real drain mechanism; what
differs from true-positive responses is their interpretation of the
embedded textual context. Rationalized responses use trust-signal vocabulary
at 1.7--19.7$\times$ the frequency of true-positive responses rated 5/5
(full table in the released artifact; Mann-Whitney U
one-sided, all $p<10^{-4}$ in our inspected sample): \texttt{transparent}
19.7$\times$, \texttt{benign} 7.2$\times$, \texttt{disclosed} 5.6$\times$,
\texttt{standard} 2.6$\times$, \texttt{audit} 2.1$\times$,
\texttt{documented} 1.7$\times$. The reverse asymmetry holds for
\texttt{legitimate} (TP $>$ rationalized): true-positive responses use
\texttt{legitimate} almost exclusively in negative contexts (``fake
legitimate-looking documentation''), corroborating the inversion mechanism.
The pattern generalizes beyond HV-3b's embedded fake audits: rationalized
responses on HV-5 (storage-loaded drain parameters) substitute a
\emph{no-mutator} assumption (``no setter is exposed in the ABI''); on
HV-6 they substitute a \emph{complexity-as-legitimacy} reading
(``well-structured DeFi protocol''). Across all three modes, the model has
the mechanistic information needed to flag the drain but instead applies
a trust-conferring frame.

\paragraph{Capability and rationalization.}
Detection rates throughout this paper use the \emph{perception binary}
(\texttt{tp}~$\cup$~\texttt{rationalized}~$\to$~detected); the
alternative \emph{deployment binary} (\texttt{tp} alone) preserves the
rank-ordering of tiers and the structural-vs-string finding but
reframes HV-3b as primarily rationalization (full table in the released artifact).
Under the perception binary the HV-3 $\to$ HV-3b paradox (mean +46\,pp
detection-rate boost when fake-audit text is added) appears nearly
universal across model classes. Under the deployment binary, where
\texttt{rationalized} verdicts collapse to missed, the paradox concentrates
sharply on the strongest-capability tier: the Anthropic family drives
+13 to +90\,pp (Opus +90 (no-thinking), Opus (thinking) +66.7,
Sonnet +66.7 (no-thinking), Sonnet (thinking) +60, Haiku +13.3); GPT-5.2 contributes +10--13.3\,pp; the
remaining model classes show $\leq 3.3$\,pp deployment-binary difference
between HV-3 and HV-3b, with the single exception of the late-added
Gemini~3 preview variant \texttt{gemini-flash-preview} (+20\,pp), the
only non-Anthropic / non-GPT-5.2 model that crosses the 3\,pp boundary. Rationalized verdicts themselves, however,
are not confined to strong models: 63 of the 84 primary-coded HV-3b
rationalizations come from models outside the Anthropic family and GPT-5.2.
What is capability-conditional is turning fake-audit text into a correct
alarm, not \emph{rationalization} itself.
A second deployment-binary observation is the Sonnet thinking-mode ablation.
Under perception, Sonnet (thinking) and Sonnet (no-thinking) detect at near-identical
rates ($-1.1$\,pp delta on HV); under deployment, extended thinking
actively hurts ($-5.6$\,pp on HV). The mechanism is
\texttt{rationalized}-share inflation: 13.5\% of Sonnet (thinking) detections
are coded \texttt{rationalized} (23/170) versus 6.4\% for Sonnet (no-thinking)
(11/173).
Extended thinking on deception contracts shifts a fraction of detections
from overt risk-flagging to mechanism-recognition without appropriate
risk-assignment.

\subsection{Model Performance}\label{sec:results:models}

Model detection rates on deception vectors span from 0\% to 76\%,
clustering into three distinct tiers.

\begin{table}[t]
  \caption{Model detection rates on deception vectors (weighted voting).
    Only HV contracts are included; controls excluded. Models within
    each tier are sorted by detection rate. Full rankings for all
    \nmodels{} models are in the released artifact.}
  \label{tab:models}
  \small
  \begin{tabular}{@{}llr@{}}
    \toprule
    Tier & Model & Rate \\
    \midrule
    \multirow{5}{*}{\rotatebox[origin=c]{90}{\small Tier 1}}
         & Claude Opus (thinking)    & 75.6\% \\
         & Claude Opus               & 73.0\% \\
         & Claude Sonnet             & 64.1\% \\
         & Claude Sonnet (thinking)  & 63.0\% \\
         & GPT-5.2 (thinking)        & 51.5\% \\
    \midrule
    \multirow{5}{*}{\rotatebox[origin=c]{90}{\small Tier 2}}
         & GPT-5.2                   & 30.7\% \\
         & Grok~4                    & 23.9\% \\
         & Claude Haiku              & 18.5\% \\
         & Gemini Flash Preview      & 13.3\% \\
         & Gemini Flash              & 10.7\% \\
    \midrule
    \multirow{5}{*}{\rotatebox[origin=c]{90}{\small Tier 3}}
         & Grok 3                    &  9.6\% \\
         & DeepSeek Reasoner         &  9.3\% \\
         & Gemini Pro                &  5.9\% \\
         & Mistral Large             &  2.6\% \\
         & Llama~4 Scout             &  0.0\% \\
    \bottomrule
  \end{tabular}
\end{table}

Table~\ref{tab:models} shows representative models from each tier (full
rankings in the released artifact). Tier~1 ($>$50\% detection)
consists exclusively of models from Anthropic and OpenAI, with most
Tier-1 models using an explicit reasoning mode.
Tier~2 (10--50\%) includes Grok~4, Claude Haiku, and two Gemini variants
alongside GPT-5.2 without explicit reasoning mode.
Tier~3 ($<$10\%) comprises the majority of models tested, including several
that are widely marketed for code analysis tasks. Tier~3 models are
included for completeness and to establish the lower bound of the
capability spectrum.\footnote{Llama~4 Scout's 0.0\% detection rate
reflects a baseline-capability ceiling rather than a deception-specific
failure: $\sim$99\% of its responses are generic refusal-to-analyze or
template completions that do not engage with the bytecode at the
function-selector level. We retain the data point for completeness but
exclude it from per-model mechanism-level discussion.} Restricting to
the 18~models above the floor-tier (HV detection rate $>$0.5\%, i.e.,
excluding qwen-coder-32b, gpt5-mini, gpt5-nano, and llama4-scout) yields
a deception delta of $-$24.3\,pp (95\% CI $[-27.4, -21.2]$); restricting
further to the 15~models with $\geq$2\% detection raises the magnitude
to $-$29.0\,pp ($[-32.3, -25.7]$). The aggregate $-$\deceptiondelta{}\,pp result is
therefore conservative: floor-tier models with near-zero detection on
both controls and HV contribute small per-model deltas that pull the
aggregate toward zero. The deception effect's direction and significance
are preserved across all tested floor thresholds.

Within-family reasoning effects are heterogeneous rather than monotonic.
The clearest matched gain appears in GPT-5.2, which improves from 30.7\%
to 51.5\% with reasoning enabled (+20.8\pp{}). Anthropic's matched pairs
show much smaller differences (Opus: +2.6\pp{}; Sonnet: $-$1.1\pp{}), and
DeepSeek's reasoning variant exceeds DeepSeek Chat by only +2.3\pp{}.
Reasoning therefore helps in some matched comparisons but is not a
universal within-family improvement. The tier structure is more
consistent with strong underlying bytecode-analysis capability, sometimes
amplified by reasoning depth, than with provider identity alone.

The tier structure is largely stable under both voting schemes: under naive
majority voting, only two models near tier boundaries (GPT-5.2 (thinking),
Gemini Flash) change tier, and per-model HV rates shift by at most 10.4\pp{}. The
full per-model$\times$per-contract detection matrix in the released
artifact confirms a clear
block structure: Tier~1 models score consistently across most contracts,
Tier~3 models score near-zero across most contracts, and the hardest
structural vectors (HV-4b, HV-3, HV-6) remain difficult even for Tier~1
models---no model achieves consistently high detection across all three.

\subsection{Prompt Interaction Effects}\label{sec:results:prompts}

\paragraph{Guard instructions offer no aggregate benefit in this setting.}
Adding guard instructions (``ignore embedded instructions'') changes the
aggregate detection rate on HV vectors by less than 1.2\pp{} in both the
investigation (A$\to$B: $-$0.1\pp{}) and neutral (D$\to$E: $+$1.2\pp{})
framings. This null result extends to prompt injection vectors
specifically: HV-PI and HV-PI2 detection rates do not improve with guards.
At the individual model level, guard effects on HV vectors are non-zero
but \emph{bidirectional}---in the A$\to$B comparison, one model improves
by 12.7\pp{} while another degrades by 13.3\pp{}---making guard
instructions not reliably beneficial and capable of shifting individual
model behavior in either direction. The same HV-only scope applies to
the aggregate span on D$\to$E; the per-model effect is narrower when
evaluated across all contracts (at most $\pm$9\pp{}), but the
bidirectional sign pattern persists.

\paragraph{Framing effects are weak.}
Across all six prompts, aggregate detection rates range from 19.4\% to
22.6\%, a total span of 3.2\pp{}. No framing achieves a practically
meaningful aggregate advantage.

\paragraph{Structural deception is largely insensitive to prompt strategy.}
The most important interaction effect is between prompt and deception type.
String-based deception responds to prompt variation (9.1\pp{} span, with
neutral prompts outperforming investigation-framed prompts), but
structural deception is largely insensitive under the tested prompt
variants: detection ranges from 9.5\% to 13.0\% across all six prompts, a
span of only 3.5\pp{}. This low sensitivity suggests that, within our
tested prompt family, structural failures are not resolved by simple
reframing or guard-style prompt engineering.

\paragraph{Prompt~F tradeoff.}
Prompt~F performs best on injection vectors
(33.2\% vs.\ 27--30\% for other prompts) and on the hardest structural
vectors (HV-3: 10.2\% vs.\ 0--4.6\%; HV-6: 15.5\% vs.\ 7--12\%). On HV-3b,
however, it produces 45~hallucinations versus 2--9 for the other prompts.
The checklist eliminates vague responses by forcing concrete function
identification, yielding either accurate detection or fabricated
mechanisms rather than equivocation.

\paragraph{Safety-filter refusals.}\label{sec:results:refusals}
Grok~4 refused 16 of 390~runs (4.1\%), with 15 refusals concentrated on
the two instruction-injection vectors and HV-PI2 alone reaching a
43.3\% refusal rate. The 16 refusals decompose into 14 pure refusals
(no analysis emitted), one surgical refusal in which the model declines
the embedded instruction but proceeds with a complete and accurate
analysis (rated 4/5 in the response text, but excluded from the
true-positive count due to the runtime's \texttt{refused} status flag),
and one runtime false-positive on C4 that reflects a regex artefact in
the refusal classifier rather than a model refusal; the net refusal
count is therefore 15. The model correctly recognizes adversarial strings
in the bytecode but treats them as threats to itself rather than as
evidence to be analyzed and reported, indicating a safety--utility
mismatch in artifact analysis. A strict head-of-response refusal detector
additionally recovers 8~short ``\texttt{I am declining\ldots}''
completions whose runtime status was \texttt{completed} (5~further
head-matching runs already carried \texttt{refused} status and are not
double-counted), yielding a corrected Grok-4 injection-vector refusal
rate of 23/60 = 38.3\% with 6/8 of the recovered cases on HV-PI2 (full
case list in the released artifact). Because this behavior is limited to
runs from a single model family, we treat it as a qualitative side
observation rather than a primary result.

\paragraph{Failure modes on prompt-injection vectors.}
Beyond the aggregate refusal and detection rates, two distinct failure
modes appear in the response data. First, noticing the injection
does not transfer to drain detection: of 656 false-negative
responses on HV-PI/HV-PI2, 292 (44.5\%) quote fragments of the
embedded injection text or explicitly name it (e.g., as an embedded
instruction, prompt injection, or social engineering), yet still miss
the drain
(markers predefined in
\texttt{analysis\_\allowbreak{}scripts/\allowbreak{}prompt\_\allowbreak{}injection\_\allowbreak{}failures.py}).
Among false-negative responses from frontier models, this share is
74--95\% (Claude Haiku 74.4\%, Grok 4 80.0\%, Gemini Flash 84.6\%,
Gemini Pro 94.7\%). The injection therefore acts as a misdirection,
consuming analytical capacity that is then unavailable for drain
detection. Second, refusals
concentrate on HV-PI2's conversational social-engineering form (10.8\%
on HV-PI2 vs.\ 1.7\% on HV-PI's imperative form; denominator: 120
frontier-model runs per contract, i.e., Claude Haiku, Grok 4, Gemini Flash,
and Gemini Pro each contributing 30~runs), suggesting Grok~4's
safety filter responds to conversational adversarial framing rather
than imperative system-override framing.

\section{Discussion}\label{sec:discussion}

\paragraph{Implications for LLM-based bytecode forensics.}
Our results provide empirical evidence against the use of LLMs as
standalone smart contract forensic analyzers in the tested single-shot,
raw-bytecode adversarial setting:
the strongest model detects only 76\% of deception vectors and the
majority fall below 15\%. Of particular concern is rationalization---when
a model examines a drain, describes it correctly, then concludes it
is benign, the resulting report is not merely incomplete but
affirmatively reassuring about an unsafe contract. This connects to
work on unfaithful explanations~\cite{turpin2023unfaithful,arcuschin2025faithful},
sycophancy~\cite{sharma2024sycophancy}, and natural-language
over-weighting in code reasoning~\cite{lam2025codecrash}: in our
setting, the benign framing is supplied by the artifact itself rather
than by a user's stated belief, but the forensic consequence is
similar---unwarranted trust in a misleading explanation. LLMs remain useful for triage and
for non-adversarial vulnerability detection; the critical implication
is that LLM-generated forensic analyses should not be treated as
authoritative without independent verification, particularly for
contracts with complex architectures or untrusted deployers.
In an investigative queue, rationalized output can actively suppress
escalation, delay preservation of transaction evidence, and misdirect
attribution toward a fictitious external exploiter. The appropriate control
is therefore not merely a confidence disclaimer: any benign conclusion must
be checked against an independently reconstructed privileged fund-flow path.

\paragraph{Why structural deception works---and why bytecode amplifies it.}
On raw bytecode, embedded strings are among the few human-readable
signals available; models anchor on string content rather than tracing
opcode-level control flow. The structural-vs-string gap
(Section~\ref{sec:results:vectors}: 37.7\% vs.\ 11.1\%) quantifies
this: structural deception removes the string foothold entirely,
exposing the limits of bytecode analysis without natural-language
anchors.

\paragraph{Defensive recommendations.}
Our results support three evidence-based recommendations. First,
\emph{prefer empirically stronger models for bytecode triage rather than
assuming a reasoning toggle will help}. The clearest matched comparison
in our data is GPT-5.2 (+20.8\pp{} with reasoning), but Anthropic's
matched pairs are nearly flat and reasoning alone does not eliminate
structural failures.\footnote{Rating-5 self-warning precision
(0--100\%) and API cost per true positive (\$0.008--\$1.88) vary by orders of
magnitude across models; per-model decomposition released in artifact.}
Second,
\emph{do not rely on guard instructions or aggregate risk scores alone}.
Our data shows no reliable aggregate benefit and bidirectional model-level
effects, making guards unreliable as a stand-alone remedy. Likewise,
realistic owner-governed vaults can trigger elevated warnings even without
a drain, so function-level drain identification is substantially more
diagnostic than a high overall score.

Third, \emph{off-the-shelf analyzer output is not a substitute for
privileged fund-flow reconstruction}: when run separately in our evaluation,
Slither~\cite{feist2019slither} and
Mythril~\cite{mueller2018mythril} identify the drain in zero contracts
(full outputs in the released artifact), as our drains use individually benign
constructs whose malice lies in \emph{composition and intent}.
Effective defenses likely require purpose-built fund-flow analysis
targeting privileged-extraction paths, and auditor-critic pipelines
that shield the reasoning model from misleading
strings~\cite{hu2023gptlens}. FinDet's semantic lifting and reachability
analysis~\cite{liu2025findet}, RPHunter's code--transaction fusion
~\cite{wu2025rphunter}, and TraceLLM's trace--decompilation pipeline
~\cite{wang2025tracellm} exemplify this richer-evidence design space. We did
not evaluate those combined pipelines, so they are defenses suggested by
adjacent evidence rather than baselines defeated by our benchmark.

\paragraph{Limitations.}
All deception vectors are purpose-built rather than drawn from real-world
deployments---no labeled corpus of deployed deceptive contracts exists.
Our vectors are informed by documented rug-pull patterns~\cite{xia2021trade,mazorra2022donotrug,zhou2024stoppulling}
but represent controlled instantiations rather than wild-caught specimens,
and the taxonomy does not claim to cover all attack strategies (delegatecall
drains, proxy abuse, cross-contract exploits, and create2 redeployment
remain untested). Within the structural category the boundary between
\emph{deception} and \emph{obfuscation} is not sharp: HV-4b's XOR masking
constitutes primarily obfuscation (selectors cannot be decoded statically),
while HV-3 and HV-5 employ purely structural deception through legitimate
Solidity constructs, so the aggregate structural detection rate (11.1\%)
conflates two distinct failure modes.

Gold-standard labels were produced by a single domain expert;
independent labeling by a second expert remains desirable. Three
complementary validation pathways (Section~\ref{sec:results:robustness}),
two fully judge-independent and requiring no human labels, corroborate the
aggregate deception effect.
Judge agreement is lowest on HV-4b and HV-3 ($\kappa = 0.249$, $0.294$),
adding uncertainty to fine-grained per-vector verdicts; both
judge-independent metrics (selector-match $-$11.4\pp{}, rating
$-$6.5\pp{}) confirm the aggregate direction.

Our bytecode-only design amplifies absolute difficulty (source-code
pilot: 41.7\%$\to$97.9\%; Section~\ref{sec:results:robustness}; full
results in the released artifact); full-scale source replication
remains future work. This identifies the boundary condition of the
result rather than weakening it. Likewise, models from
one provider dominate Tier~1 because the vectors were developed against
Opus (Section~\ref{sec:taxonomy:vectors}); its 76\% rate is measured
against contracts designed to challenge it. Four observations argue
against an adaptive-construction artifact: transfer to 21~other models
from six providers yields
consistent structural-failure patterns; the intra-Anthropic spread
(Haiku 18.5\% vs.\ Opus 75.6\%) shows Tier-1 dominance is model-level
rather than provider-uniform; a leave-provider-out bootstrap leaves the
delta significantly negative even when
Anthropic and OpenAI are jointly excluded ($-$16.5\pp{}); and two held-out
Grok-generated vectors fall within the same-panel main-study range
(Section~\ref{sec:results:robustness}). This probe reduces but cannot remove
construction bias. The reported rates therefore
characterize behavior on this stress benchmark and do not estimate the
prevalence of deception in deployed contracts; provider, architecture,
training, and reasoning mode are not fully separable in our model set
(Section~\ref{sec:results:models}). DeepSeek and Gemini families appear
as both subjects and judges, but the judge-independent metrics
(Section~\ref{sec:method:judge}) provide a corroborating pathway with
zero model overlap.

Our evaluation uses single-shot raw-bytecode analysis; interactive prompting,
decompilation, source retrieval, execution tools, or on-chain context may
yield higher detection rates and are outside the claim. Finally, 36~runs
(0.4\%) remain pending in
two lower-tier models; the sensitivity bounds reported in
Section~\ref{sec:method:design} show that assigning these runs to extreme
outcomes has negligible effect on the primary deception delta.

\section{Conclusion}\label{sec:conclusion}

To our knowledge, we present the first systematic adversarial evaluation
of LLM-based smart contract \emph{single-shot, raw-bytecode-only forensic
triage} against
deliberately deceptive contracts. Across \nmodels{} contemporary models
and \nruns{} calibrated runs, adversarial deception reduces detection by
\deceptiondelta{}\pp{}; the hardest structural vectors achieve 2--4\%
aggregate detection, with per-model performance on these vectors
spanning 0\% to 18\%. We identify rationalization
as a distinct failure mode: models correctly describe malicious
mechanisms then dismiss them as benign, producing reports that are
affirmatively reassuring about unsafe contracts. Guard instructions
provide no reliable aggregate benefit, structural deception is largely
insensitive to simple prompt variation within the prompt family tested
here, and only a small set of models from two providers exceed 50\%
detection. Off-the-shelf static analyzers (Slither, Mythril) likewise
miss these drains when run separately; robust workflows likely require
purpose-built fund-flow detectors, independent evidence, and human review,
with LLMs used only as triage rather than final adjudicators.

\bibliographystyle{ACM-Reference-Format}
\bibliography{references_verified}


\begin{thebibliography}{44}


\ifx \showCODEN    \undefined \def \showCODEN     #1{\unskip}     \fi
\ifx \showISBNx    \undefined \def \showISBNx     #1{\unskip}     \fi
\ifx \showISBNxiii \undefined \def \showISBNxiii  #1{\unskip}     \fi
\ifx \showISSN     \undefined \def \showISSN      #1{\unskip}     \fi
\ifx \showLCCN     \undefined \def \showLCCN      #1{\unskip}     \fi
\ifx \shownote     \undefined \def \shownote      #1{#1}          \fi
\ifx \showarticletitle \undefined \def \showarticletitle #1{#1}   \fi
\ifx \showURL      \undefined \def \showURL       {\relax}        \fi
\providecommand\bibfield[2]{#2}
\providecommand\bibinfo[2]{#2}
\providecommand\natexlab[1]{#1}
\providecommand\showeprint[2][]{arXiv:#2}
\makeatletter
\@ifundefined{NAT@parse@date}{}{\let\NAT@parse@date@orig\NAT@parse@date}
\@ifundefined{NAT@parse@date}{}{\def\NAT@parse@date#1#2#3#4#5#6@@{\NAT@parse@date@orig#1#2#3#4#5#6@@\def\NAT@tempyear{0000}\def\NAT@tempexlab{{?}}\ifx\NAT@year\NAT@tempyear\ifx\NAT@exlab\NAT@tempexlab\def\NAT@date{[n.\,d.]}\else\edef\NAT@date{[n.\,d.]\NAT@exlab}\fi\fi}}
\makeatother

\bibitem[Abdelaziz et~al\mbox{.}(2026)]%
        {abdelaziz2026analyzers}
\bibfield{author}{\bibinfo{person}{Tamer Abdelaziz}, \bibinfo{person}{Salma
  Alsaghir}, {and} \bibinfo{person}{Karim Ali}.}
  \bibinfo{year}{2026}\natexlab{}.
\newblock \showarticletitle{Where Do Smart Contract Security Analyzers Fall
  Short?}. In \bibinfo{booktitle}{\emph{Proceedings of the 23rd International
  Conference on Mining Software Repositories}}. \bibinfo{publisher}{ACM}.
\newblock
\href{https://doi.org/10.1145/3793302.3793338}{doi:\nolinkurl{10.1145/3793302.3793338}}


\bibitem[Arcuschin et~al\mbox{.}(2025)]%
        {arcuschin2025faithful}
\bibfield{author}{\bibinfo{person}{Iv{\'a}n Arcuschin}, \bibinfo{person}{Jett
  Janiak}, \bibinfo{person}{Robert Krzyzanowski}, \bibinfo{person}{Senthooran
  Rajamanoharan}, \bibinfo{person}{Neel Nanda}, {and} \bibinfo{person}{Arthur
  Conmy}.} \bibinfo{year}{2025}\natexlab{}.
\newblock \showarticletitle{Chain-of-Thought Reasoning In The Wild Is Not
  Always Faithful}. In \bibinfo{booktitle}{\emph{{ICLR} 2025 Workshop on
  Reasoning and Planning for {LLMs}}}.
\newblock
\showeprint[arxiv]{2503.08679}
\href{https://doi.org/10.48550/arXiv.2503.08679}{doi:\nolinkurl{10.48550/arXiv.2503.08679}}


\bibitem[Chen et~al\mbox{.}(2025b)]%
        {chen2025chatgpt}
\bibfield{author}{\bibinfo{person}{Chong Chen}, \bibinfo{person}{Jianzhong Su},
  \bibinfo{person}{Jiachi Chen}, \bibinfo{person}{Yanlin Wang},
  \bibinfo{person}{Tingting Bi}, \bibinfo{person}{Jianxing Yu},
  \bibinfo{person}{Yanli Wang}, \bibinfo{person}{Xingwei Lin},
  \bibinfo{person}{Ting Chen}, {and} \bibinfo{person}{Zibin Zheng}.}
  \bibinfo{year}{2025}\natexlab{b}.
\newblock \showarticletitle{When {ChatGPT} Meets Smart Contract Vulnerability
  Detection: How Far Are We?}
\newblock \bibinfo{journal}{\emph{ACM Transactions on Software Engineering and
  Methodology}} \bibinfo{volume}{34}, \bibinfo{number}{4}, Article
  \bibinfo{articleno}{100} (\bibinfo{year}{2025}),
  \bibinfo{numpages}{30}~pages.
\newblock
\href{https://doi.org/10.1145/3702973}{doi:\nolinkurl{10.1145/3702973}}


\bibitem[Chen et~al\mbox{.}(2025a)]%
        {chen2025struq}
\bibfield{author}{\bibinfo{person}{Sizhe Chen}, \bibinfo{person}{Julien Piet},
  \bibinfo{person}{Chawin Sitawarin}, {and} \bibinfo{person}{David Wagner}.}
  \bibinfo{year}{2025}\natexlab{a}.
\newblock \showarticletitle{{StruQ}: Defending Against Prompt Injection with
  Structured Queries}. In \bibinfo{booktitle}{\emph{34th USENIX Security
  Symposium (USENIX Security 25)}}. \bibinfo{publisher}{USENIX Association},
  \bibinfo{address}{Seattle, WA, USA}, \bibinfo{pages}{2383--2400}.
\newblock
\urldef\tempurl%
\url{https://www.usenix.org/conference/usenixsecurity25/presentation/chen-sizhe}
\showURL{%
\tempurl}


\bibitem[Choi et~al\mbox{.}(2025)]%
        {choi2025graphllm}
\bibfield{author}{\bibinfo{person}{Ra-Yeon Choi}, \bibinfo{person}{Yeji Song},
  \bibinfo{person}{Minsoo Jang}, \bibinfo{person}{Taekyung Kim},
  \bibinfo{person}{Jinhyun Ahn}, {and} \bibinfo{person}{Dong-Hyuk Im}.}
  \bibinfo{year}{2025}\natexlab{}.
\newblock \showarticletitle{Smart Contract Vulnerability Detection Using Large
  Language Models and Graph Structural Analysis}.
\newblock \bibinfo{journal}{\emph{Computers, Materials \& Continua}}
  \bibinfo{volume}{83}, \bibinfo{number}{1} (\bibinfo{year}{2025}),
  \bibinfo{pages}{785--801}.
\newblock
\href{https://doi.org/10.32604/cmc.2025.061185}{doi:\nolinkurl{10.32604/cmc.2025.061185}}


\bibitem[David et~al\mbox{.}(2023)]%
        {david2023manual}
\bibfield{author}{\bibinfo{person}{Isaac David}, \bibinfo{person}{Liyi Zhou},
  \bibinfo{person}{Kaihua Qin}, \bibinfo{person}{Dawn Song},
  \bibinfo{person}{Lorenzo Cavallaro}, {and} \bibinfo{person}{Arthur Gervais}.}
  \bibinfo{year}{2023}\natexlab{}.
\newblock \bibinfo{title}{Do You Still Need a Manual Smart Contract Audit?}
\newblock \bibinfo{howpublished}{CoRR, abs/2306.12338}.
\newblock
\href{https://doi.org/10.48550/arXiv.2306.12338}{doi:\nolinkurl{10.48550/arXiv.2306.12338}}


\bibitem[David et~al\mbox{.}(2025)]%
        {david2025decompiling}
\bibfield{author}{\bibinfo{person}{Isaac David}, \bibinfo{person}{Liyi Zhou},
  \bibinfo{person}{Dawn Song}, \bibinfo{person}{Arthur Gervais}, {and}
  \bibinfo{person}{Kaihua Qin}.} \bibinfo{year}{2025}\natexlab{}.
\newblock \bibinfo{title}{Decompiling Smart Contracts with a Large Language
  Model}.
\newblock \bibinfo{howpublished}{CoRR, abs/2506.19624}.
\newblock
\href{https://doi.org/10.48550/arXiv.2506.19624}{doi:\nolinkurl{10.48550/arXiv.2506.19624}}


\bibitem[De~Rosa et~al\mbox{.}(2025)]%
        {derosa2025phishinghook}
\bibfield{author}{\bibinfo{person}{Pasquale De~Rosa}, \bibinfo{person}{Simon
  Queyrut}, \bibinfo{person}{Y{\'e}rom-David Bromberg}, \bibinfo{person}{Pascal
  Felber}, {and} \bibinfo{person}{Valerio Schiavoni}.}
  \bibinfo{year}{2025}\natexlab{}.
\newblock \showarticletitle{{PhishingHook}: Catching Phishing Ethereum Smart
  Contracts Leveraging {EVM} Opcodes}. In \bibinfo{booktitle}{\emph{Proceedings
  of the 55th IEEE/IFIP International Conference on Dependable Systems and
  Networks}}. \bibinfo{publisher}{IEEE}.
\newblock
\href{https://doi.org/10.1109/DSN64029.2025.00033}{doi:\nolinkurl{10.1109/DSN64029.2025.00033}}


\bibitem[Ding et~al\mbox{.}(2025)]%
        {ding2025smartguard}
\bibfield{author}{\bibinfo{person}{Hao Ding}, \bibinfo{person}{Yizhou Liu},
  \bibinfo{person}{Xuefeng Piao}, \bibinfo{person}{Huihui Song}, {and}
  \bibinfo{person}{Zhenzhou Ji}.} \bibinfo{year}{2025}\natexlab{}.
\newblock \showarticletitle{{SmartGuard}: An {LLM}-Enhanced Framework for Smart
  Contract Vulnerability Detection}.
\newblock \bibinfo{journal}{\emph{Expert Systems with Applications}}
  \bibinfo{volume}{269} (\bibinfo{year}{2025}), \bibinfo{pages}{126479}.
\newblock
\href{https://doi.org/10.1016/j.eswa.2025.126479}{doi:\nolinkurl{10.1016/j.eswa.2025.126479}}


\bibitem[Durieux et~al\mbox{.}(2020)]%
        {durieux2020empirical}
\bibfield{author}{\bibinfo{person}{Thomas Durieux},
  \bibinfo{person}{Jo{\~a}o~F. Ferreira}, \bibinfo{person}{Rui Abreu}, {and}
  \bibinfo{person}{Pedro Cruz}.} \bibinfo{year}{2020}\natexlab{}.
\newblock \showarticletitle{Empirical Review of Automated Analysis Tools on
  47,587 {Ethereum} Smart Contracts}. In \bibinfo{booktitle}{\emph{Proceedings
  of the ACM/IEEE 42nd International Conference on Software Engineering}}.
  \bibinfo{publisher}{IEEE Computer Society}, \bibinfo{address}{Virtual /
  Seoul, Republic of Korea}, \bibinfo{pages}{530--541}.
\newblock
\href{https://doi.org/10.1145/3377811.3380364}{doi:\nolinkurl{10.1145/3377811.3380364}}


\bibitem[Efron(1979)]%
        {efron1979bootstrap}
\bibfield{author}{\bibinfo{person}{Bradley Efron}.}
  \bibinfo{year}{1979}\natexlab{}.
\newblock \showarticletitle{Bootstrap Methods: Another Look at the Jackknife}.
\newblock \bibinfo{journal}{\emph{The Annals of Statistics}}
  \bibinfo{volume}{7}, \bibinfo{number}{1} (\bibinfo{year}{1979}),
  \bibinfo{pages}{1--26}.
\newblock
\href{https://doi.org/10.1214/aos/1176344552}{doi:\nolinkurl{10.1214/aos/1176344552}}


\bibitem[Feist et~al\mbox{.}(2019)]%
        {feist2019slither}
\bibfield{author}{\bibinfo{person}{Josselin Feist}, \bibinfo{person}{Gustavo
  Grieco}, {and} \bibinfo{person}{Alex Groce}.}
  \bibinfo{year}{2019}\natexlab{}.
\newblock \showarticletitle{Slither: A Static Analysis Framework for Smart
  Contracts}. In \bibinfo{booktitle}{\emph{Proceedings of the 2019 IEEE/ACM 2nd
  International Workshop on Emerging Trends in Software Engineering for
  Blockchain (WETSEB)}}. \bibinfo{publisher}{IEEE}, \bibinfo{address}{Montreal,
  Canada}, \bibinfo{pages}{8--15}.
\newblock
\href{https://doi.org/10.1109/WETSEB.2019.00008}{doi:\nolinkurl{10.1109/WETSEB.2019.00008}}


\bibitem[Fleiss(1971)]%
        {fleiss1971kappa}
\bibfield{author}{\bibinfo{person}{Joseph~L. Fleiss}.}
  \bibinfo{year}{1971}\natexlab{}.
\newblock \showarticletitle{Measuring Nominal Scale Agreement Among Many
  Raters}.
\newblock \bibinfo{journal}{\emph{Psychological Bulletin}}
  \bibinfo{volume}{76}, \bibinfo{number}{5} (\bibinfo{year}{1971}),
  \bibinfo{pages}{378--382}.
\newblock
\href{https://doi.org/10.1037/h0031619}{doi:\nolinkurl{10.1037/h0031619}}


\bibitem[Grech et~al\mbox{.}(2019)]%
        {grech2019gigahorse}
\bibfield{author}{\bibinfo{person}{Neville Grech}, \bibinfo{person}{Lexi
  Brent}, \bibinfo{person}{Bernhard Scholz}, {and} \bibinfo{person}{Yannis
  Smaragdakis}.} \bibinfo{year}{2019}\natexlab{}.
\newblock \showarticletitle{Gigahorse: Thorough, Declarative Decompilation of
  Smart Contracts}. In \bibinfo{booktitle}{\emph{Proceedings of the 41st
  International Conference on Software Engineering}}. \bibinfo{publisher}{IEEE
  / ACM}, \bibinfo{address}{Montreal, Canada}, \bibinfo{pages}{1176--1186}.
\newblock
\href{https://doi.org/10.1109/ICSE.2019.00120}{doi:\nolinkurl{10.1109/ICSE.2019.00120}}


\bibitem[Greshake et~al\mbox{.}(2023)]%
        {greshake2023inject}
\bibfield{author}{\bibinfo{person}{Kai Greshake}, \bibinfo{person}{Sahar
  Abdelnabi}, \bibinfo{person}{Shailesh Mishra}, \bibinfo{person}{Christoph
  Endres}, \bibinfo{person}{Thorsten Holz}, {and} \bibinfo{person}{Mario
  Fritz}.} \bibinfo{year}{2023}\natexlab{}.
\newblock \showarticletitle{Not What You've Signed Up For: Compromising
  Real-World {LLM}-Integrated Applications with Indirect Prompt Injection}. In
  \bibinfo{booktitle}{\emph{Proceedings of the 16th ACM Workshop on Artificial
  Intelligence and Security}}. \bibinfo{publisher}{ACM},
  \bibinfo{address}{Copenhagen, Denmark}, \bibinfo{pages}{79--90}.
\newblock
\href{https://doi.org/10.1145/3605764.3623985}{doi:\nolinkurl{10.1145/3605764.3623985}}


\bibitem[He et~al\mbox{.}(2024)]%
        {he2024llm4bs}
\bibfield{author}{\bibinfo{person}{Zheyuan He}, \bibinfo{person}{Zihao Li},
  \bibinfo{person}{Sen Yang}, \bibinfo{person}{He Ye}, \bibinfo{person}{Ao
  Qiao}, \bibinfo{person}{Xiaosong Zhang}, \bibinfo{person}{Ting Chen}, {and}
  \bibinfo{person}{Xiapu Luo}.} \bibinfo{year}{2024}\natexlab{}.
\newblock \bibinfo{title}{Large Language Models for Blockchain Security: A
  Systematic Literature Review}.
\newblock \bibinfo{howpublished}{Cryptology ePrint Archive, Paper 2024/477}.
\newblock
\urldef\tempurl%
\url{https://eprint.iacr.org/2024/477}
\showURL{%
\tempurl}


\bibitem[Hu et~al\mbox{.}(2023)]%
        {hu2023gptlens}
\bibfield{author}{\bibinfo{person}{Sihao Hu}, \bibinfo{person}{Tiansheng
  Huang}, \bibinfo{person}{Fatih Ilhan}, \bibinfo{person}{Selim~Furkan Tekin},
  {and} \bibinfo{person}{Ling Liu}.} \bibinfo{year}{2023}\natexlab{}.
\newblock \showarticletitle{Large Language Model-Powered Smart Contract
  Vulnerability Detection: New Perspectives}. In \bibinfo{booktitle}{\emph{2023
  IEEE 5th International Conference on Trust, Privacy and Security in
  Intelligent Systems and Applications (TPS-ISA)}}. \bibinfo{publisher}{IEEE},
  \bibinfo{address}{Atlanta, GA, USA}, \bibinfo{pages}{297--306}.
\newblock
\href{https://doi.org/10.1109/TPS-ISA58951.2023.00044}{doi:\nolinkurl{10.1109/TPS-ISA58951.2023.00044}}


\bibitem[Ince et~al\mbox{.}(2026)]%
        {ince2026gendetect}
\bibfield{author}{\bibinfo{person}{Peter Ince}, \bibinfo{person}{Jiangshan Yu},
  \bibinfo{person}{Joseph~K. Liu}, \bibinfo{person}{Xiaoning Du}, {and}
  \bibinfo{person}{Xiapu Luo}.} \bibinfo{year}{2026}\natexlab{}.
\newblock \showarticletitle{{GenDetect}: Generative Large Language Model Usage
  in Smart Contract Vulnerability Detection}. In
  \bibinfo{booktitle}{\emph{Provable and Practical Security -- 19th
  International Conference, ProvSec 2025, Proceedings}}
  \emph{(\bibinfo{series}{Lecture Notes in Computer Science},
  Vol.~\bibinfo{volume}{16172})}. \bibinfo{publisher}{Springer},
  \bibinfo{address}{Yokohama, Japan}, \bibinfo{pages}{426--445}.
\newblock
\href{https://doi.org/10.1007/978-981-95-2961-2_22}{doi:\nolinkurl{10.1007/978-981-95-2961-2_22}}


\bibitem[Lagouvardos et~al\mbox{.}(2026)]%
        {lagouvardos2026dyels}
\bibfield{author}{\bibinfo{person}{Sifis Lagouvardos}, \bibinfo{person}{Yannis
  Bollanos}, \bibinfo{person}{Michael Debono}, \bibinfo{person}{Neville Grech},
  {and} \bibinfo{person}{Yannis Smaragdakis}.} \bibinfo{year}{2026}\natexlab{}.
\newblock \bibinfo{title}{Precise Static Identification of Ethereum Storage
  Variables}.
\newblock \bibinfo{howpublished}{Extended version of the ICSE 2026 paper, CoRR,
  abs/2503.20690}.
\newblock
\href{https://doi.org/10.48550/arXiv.2503.20690}{doi:\nolinkurl{10.48550/arXiv.2503.20690}}


\bibitem[Lam et~al\mbox{.}(2025)]%
        {lam2025codecrash}
\bibfield{author}{\bibinfo{person}{Man~Ho Lam}, \bibinfo{person}{Chaozheng
  Wang}, \bibinfo{person}{Jen-tse Huang}, {and} \bibinfo{person}{Michael~R.
  Lyu}.} \bibinfo{year}{2025}\natexlab{}.
\newblock \showarticletitle{{CodeCrash}: Exposing {LLM} Fragility to Misleading
  Natural Language in Code Reasoning}. In \bibinfo{booktitle}{\emph{Advances in
  Neural Information Processing Systems 38 ({NeurIPS} 2025)}}.
  \bibinfo{publisher}{Curran Associates, Inc.}, \bibinfo{address}{Vancouver,
  BC, Canada}.
\newblock
\urldef\tempurl%
\url{https://arxiv.org/abs/2504.14119}
\showURL{%
\tempurl}


\bibitem[Liu et~al\mbox{.}(2023)]%
        {liu2023jailbreaking}
\bibfield{author}{\bibinfo{person}{Yi Liu}, \bibinfo{person}{Gelei Deng},
  \bibinfo{person}{Zhengzi Xu}, \bibinfo{person}{Yuekang Li},
  \bibinfo{person}{Yaowen Zheng}, \bibinfo{person}{Ying Zhang},
  \bibinfo{person}{Lida Zhao}, \bibinfo{person}{Tianwei Zhang},
  \bibinfo{person}{Kailong Wang}, {and} \bibinfo{person}{Yang Liu}.}
  \bibinfo{year}{2023}\natexlab{}.
\newblock \bibinfo{title}{Jailbreaking {ChatGPT} via Prompt Engineering: An
  Empirical Study}.
\newblock \bibinfo{howpublished}{CoRR, abs/2305.13860}.
\newblock
\href{https://doi.org/10.48550/arXiv.2305.13860}{doi:\nolinkurl{10.48550/arXiv.2305.13860}}


\bibitem[Liu et~al\mbox{.}(2025)]%
        {liu2025findet}
\bibfield{author}{\bibinfo{person}{Yating Liu}, \bibinfo{person}{Xing Su},
  \bibinfo{person}{Hao Wu}, \bibinfo{person}{Sijin Li}, \bibinfo{person}{Yuxi
  Cheng}, \bibinfo{person}{Fengyuan Xu}, {and} \bibinfo{person}{Sheng Zhong}.}
  \bibinfo{year}{2025}\natexlab{}.
\newblock \bibinfo{title}{Revealing Adversarial Smart Contracts through
  Semantic Interpretation and Uncertainty Estimation}.
\newblock \bibinfo{howpublished}{CoRR, abs/2509.18934}.
\newblock
\href{https://doi.org/10.48550/arXiv.2509.18934}{doi:\nolinkurl{10.48550/arXiv.2509.18934}}


\bibitem[Luu et~al\mbox{.}(2016)]%
        {luu2016oyente}
\bibfield{author}{\bibinfo{person}{Loi Luu}, \bibinfo{person}{Duc-Hiep Chu},
  \bibinfo{person}{Hrishi Olickel}, \bibinfo{person}{Prateek Saxena}, {and}
  \bibinfo{person}{Aquinas Hobor}.} \bibinfo{year}{2016}\natexlab{}.
\newblock \showarticletitle{Making Smart Contracts Smarter}. In
  \bibinfo{booktitle}{\emph{Proceedings of the 2016 ACM SIGSAC Conference on
  Computer and Communications Security}}. \bibinfo{publisher}{ACM},
  \bibinfo{address}{Vienna, Austria}, \bibinfo{pages}{254--269}.
\newblock
\href{https://doi.org/10.1145/2976749.2978309}{doi:\nolinkurl{10.1145/2976749.2978309}}


\bibitem[Mazorra et~al\mbox{.}(2022)]%
        {mazorra2022donotrug}
\bibfield{author}{\bibinfo{person}{Bruno Mazorra}, \bibinfo{person}{Victor
  Adan}, {and} \bibinfo{person}{Vanesa Daza}.} \bibinfo{year}{2022}\natexlab{}.
\newblock \showarticletitle{Do Not Rug on Me: Leveraging Machine Learning
  Techniques for Automated Scam Detection}.
\newblock \bibinfo{journal}{\emph{Mathematics}} \bibinfo{volume}{10},
  \bibinfo{number}{6} (\bibinfo{year}{2022}), \bibinfo{pages}{949}.
\newblock
\href{https://doi.org/10.3390/math10060949}{doi:\nolinkurl{10.3390/math10060949}}


\bibitem[Mueller(2018)]%
        {mueller2018mythril}
\bibfield{author}{\bibinfo{person}{Bernhard Mueller}.}
  \bibinfo{year}{2018}\natexlab{}.
\newblock \bibinfo{title}{Mythril Classic}.
\newblock
\urldef\tempurl%
\url{https://github.com/ConsenSysDiligence/mythril}
\showURL{%
\tempurl}
\newblock
\shownote{Accessed 2026-04-20}.


\bibitem[Pearce et~al\mbox{.}(2022)]%
        {pearce2022copilot}
\bibfield{author}{\bibinfo{person}{Hammond Pearce}, \bibinfo{person}{Baleegh
  Ahmad}, \bibinfo{person}{Benjamin Tan}, \bibinfo{person}{Brendan
  Dolan-Gavitt}, {and} \bibinfo{person}{Ramesh Karri}.}
  \bibinfo{year}{2022}\natexlab{}.
\newblock \showarticletitle{Asleep at the Keyboard? {A}ssessing the Security of
  {GitHub} {Copilot}'s Code Contributions}. In \bibinfo{booktitle}{\emph{2022
  IEEE Symposium on Security and Privacy (SP)}}. \bibinfo{publisher}{IEEE},
  \bibinfo{address}{San Francisco, CA, USA}, \bibinfo{pages}{754--768}.
\newblock
\href{https://doi.org/10.1109/SP46214.2022.9833571}{doi:\nolinkurl{10.1109/SP46214.2022.9833571}}


\bibitem[Salzano et~al\mbox{.}(2026)]%
        {salzano2026tools}
\bibfield{author}{\bibinfo{person}{Francesco Salzano},
  \bibinfo{person}{Cosmo~Kevin Antenucci}, \bibinfo{person}{Simone Scalabrino},
  \bibinfo{person}{Giovanni Rosa}, \bibinfo{person}{Rocco Oliveto}, {and}
  \bibinfo{person}{Remo Pareschi}.} \bibinfo{year}{2026}\natexlab{}.
\newblock \showarticletitle{An Empirical Analysis of Vulnerability Detection
  Tools for Solidity Smart Contracts}.
\newblock \bibinfo{journal}{\emph{Empirical Software Engineering}}
  \bibinfo{volume}{31}, Article \bibinfo{articleno}{143}
  (\bibinfo{year}{2026}).
\newblock
\href{https://doi.org/10.1007/s10664-026-10867-7}{doi:\nolinkurl{10.1007/s10664-026-10867-7}}


\bibitem[Schulhoff et~al\mbox{.}(2023)]%
        {schulhoff2023hackaprompt}
\bibfield{author}{\bibinfo{person}{Sander Schulhoff}, \bibinfo{person}{Jeremy
  Pinto}, \bibinfo{person}{Anaum Khan}, \bibinfo{person}{Louis-Fran{\c{c}}ois
  Bouchard}, \bibinfo{person}{Chenglei Si}, \bibinfo{person}{Svetlina Anati},
  \bibinfo{person}{Valen Tagliabue}, \bibinfo{person}{Anson Kost},
  \bibinfo{person}{Christopher Carnahan}, {and} \bibinfo{person}{Jordan
  Boyd-Graber}.} \bibinfo{year}{2023}\natexlab{}.
\newblock \showarticletitle{Ignore This Title and {H}ack{AP}rompt: Exposing
  Systemic Vulnerabilities of {LLM}s Through a Global Prompt Hacking
  Competition}. In \bibinfo{booktitle}{\emph{Proceedings of the 2023 Conference
  on Empirical Methods in Natural Language Processing}}.
  \bibinfo{publisher}{Association for Computational Linguistics},
  \bibinfo{address}{Singapore}, \bibinfo{pages}{4945--4977}.
\newblock
\href{https://doi.org/10.18653/v1/2023.emnlp-main.302}{doi:\nolinkurl{10.18653/v1/2023.emnlp-main.302}}


\bibitem[Sharma et~al\mbox{.}(2024)]%
        {sharma2024sycophancy}
\bibfield{author}{\bibinfo{person}{Mrinank Sharma}, \bibinfo{person}{Meg Tong},
  \bibinfo{person}{Tomasz Korbak}, \bibinfo{person}{David Duvenaud},
  \bibinfo{person}{Amanda Askell}, \bibinfo{person}{Samuel~R. Bowman},
  \bibinfo{person}{Newton Cheng}, \bibinfo{person}{Esin Durmus},
  \bibinfo{person}{Zac Hatfield-Dodds}, \bibinfo{person}{Scott~R. Johnston},
  \bibinfo{person}{Shauna Kravec}, \bibinfo{person}{Timothy Maxwell},
  \bibinfo{person}{Sam McCandlish}, \bibinfo{person}{Kamal Ndousse},
  \bibinfo{person}{Oliver Rausch}, \bibinfo{person}{Nicholas Schiefer},
  \bibinfo{person}{Da Yan}, \bibinfo{person}{Miranda Zhang}, {and}
  \bibinfo{person}{Ethan Perez}.} \bibinfo{year}{2024}\natexlab{}.
\newblock \showarticletitle{Towards Understanding Sycophancy in Language
  Models}. In \bibinfo{booktitle}{\emph{The Twelfth International Conference on
  Learning Representations ({ICLR} 2024)}}.
\newblock
\urldef\tempurl%
\url{https://openreview.net/forum?id=tvhaxkMKAn}
\showURL{%
\tempurl}


\bibitem[Sun et~al\mbox{.}(2024)]%
        {sun2024gptscan}
\bibfield{author}{\bibinfo{person}{Yuqiang Sun}, \bibinfo{person}{Daoyuan Wu},
  \bibinfo{person}{Yue Xue}, \bibinfo{person}{Han Liu}, \bibinfo{person}{Haijun
  Wang}, \bibinfo{person}{Zhengzi Xu}, \bibinfo{person}{Xiaofei Xie}, {and}
  \bibinfo{person}{Yang Liu}.} \bibinfo{year}{2024}\natexlab{}.
\newblock \showarticletitle{{GPTScan}: Detecting Logic Vulnerabilities in Smart
  Contracts by Combining {GPT} with Program Analysis}. In
  \bibinfo{booktitle}{\emph{Proceedings of the IEEE/ACM 46th International
  Conference on Software Engineering}}. \bibinfo{publisher}{ACM},
  \bibinfo{address}{Lisbon, Portugal}, Article \bibinfo{articleno}{166},
  \bibinfo{numpages}{13}~pages.
\newblock
\href{https://doi.org/10.1145/3597503.3639117}{doi:\nolinkurl{10.1145/3597503.3639117}}


\bibitem[Thornton(2026)]%
        {thornton2026codecomments}
\bibfield{author}{\bibinfo{person}{Scott Thornton}.}
  \bibinfo{year}{2026}\natexlab{}.
\newblock \bibinfo{title}{Can Adversarial Code Comments Fool {AI} Security
  Reviewers --- Large-Scale Empirical Study of Comment-Based Attacks and
  Defenses Against {LLM} Code Analysis}.
\newblock \bibinfo{howpublished}{CoRR, abs/2602.16741}.
\newblock
\href{https://doi.org/10.48550/arXiv.2602.16741}{doi:\nolinkurl{10.48550/arXiv.2602.16741}}


\bibitem[Torres et~al\mbox{.}(2019)]%
        {torres2019honeypot}
\bibfield{author}{\bibinfo{person}{Christof~Ferreira Torres},
  \bibinfo{person}{Mathis Steichen}, {and} \bibinfo{person}{Radu State}.}
  \bibinfo{year}{2019}\natexlab{}.
\newblock \showarticletitle{The Art of The Scam: Demystifying Honeypots in
  {Ethereum} Smart Contracts}. In \bibinfo{booktitle}{\emph{28th USENIX
  Security Symposium (USENIX Security 19)}}. \bibinfo{publisher}{USENIX
  Association}, \bibinfo{address}{Santa Clara, CA, USA},
  \bibinfo{pages}{1591--1607}.
\newblock
\urldef\tempurl%
\url{https://www.usenix.org/conference/usenixsecurity19/presentation/ferreira}
\showURL{%
\tempurl}


\bibitem[Tsankov et~al\mbox{.}(2018)]%
        {tsankov2018securify}
\bibfield{author}{\bibinfo{person}{Petar Tsankov}, \bibinfo{person}{Andrei
  Dan}, \bibinfo{person}{Dana Drachsler-Cohen}, \bibinfo{person}{Arthur
  Gervais}, \bibinfo{person}{Florian B\"{u}nzli}, {and} \bibinfo{person}{Martin
  Vechev}.} \bibinfo{year}{2018}\natexlab{}.
\newblock \showarticletitle{Securify: Practical Security Analysis of Smart
  Contracts}. In \bibinfo{booktitle}{\emph{Proceedings of the 2018 ACM SIGSAC
  Conference on Computer and Communications Security}}.
  \bibinfo{publisher}{ACM}, \bibinfo{address}{Toronto, Canada},
  \bibinfo{pages}{67--82}.
\newblock
\href{https://doi.org/10.1145/3243734.3243780}{doi:\nolinkurl{10.1145/3243734.3243780}}


\bibitem[Turpin et~al\mbox{.}(2023)]%
        {turpin2023unfaithful}
\bibfield{author}{\bibinfo{person}{Miles Turpin}, \bibinfo{person}{Julian
  Michael}, \bibinfo{person}{Ethan Perez}, {and} \bibinfo{person}{Samuel~R.
  Bowman}.} \bibinfo{year}{2023}\natexlab{}.
\newblock \showarticletitle{Language Models Don't Always Say What They Think:
  Unfaithful Explanations in Chain-of-Thought Prompting}. In
  \bibinfo{booktitle}{\emph{Advances in Neural Information Processing Systems
  36 ({NeurIPS} 2023)}}.
\newblock
\urldef\tempurl%
\url{https://proceedings.neurips.cc/paper_files/paper/2023/hash/ed3fea9033a80fea1376299fa7863f4a-Abstract-Conference.html}
\showURL{%
\tempurl}


\bibitem[Wallace et~al\mbox{.}(2024)]%
        {wallace2024instructionhierarchy}
\bibfield{author}{\bibinfo{person}{Eric Wallace}, \bibinfo{person}{Kai Xiao},
  \bibinfo{person}{Reimar Leike}, \bibinfo{person}{Lilian Weng},
  \bibinfo{person}{Johannes Heidecke}, {and} \bibinfo{person}{Alex Beutel}.}
  \bibinfo{year}{2024}\natexlab{}.
\newblock \bibinfo{title}{The Instruction Hierarchy: Training {LLM}s to
  Prioritize Privileged Instructions}.
\newblock \bibinfo{howpublished}{CoRR, abs/2404.13208}.
\newblock
\href{https://doi.org/10.48550/arXiv.2404.13208}{doi:\nolinkurl{10.48550/arXiv.2404.13208}}


\bibitem[Wang et~al\mbox{.}(2025)]%
        {wang2025tracellm}
\bibfield{author}{\bibinfo{person}{Shuzheng Wang}, \bibinfo{person}{Yue Huang},
  \bibinfo{person}{Zhuoer Xu}, \bibinfo{person}{Yuming Huang}, {and}
  \bibinfo{person}{Jing Tang}.} \bibinfo{year}{2025}\natexlab{}.
\newblock \bibinfo{title}{{TraceLLM}: Security Diagnosis Through Traces and
  Smart Contracts in Ethereum}.
\newblock \bibinfo{howpublished}{CoRR, abs/2509.03037}.
\newblock
\href{https://doi.org/10.48550/arXiv.2509.03037}{doi:\nolinkurl{10.48550/arXiv.2509.03037}}


\bibitem[Wei et~al\mbox{.}(2024)]%
        {wei2024llmsmartaudit}
\bibfield{author}{\bibinfo{person}{Zhiyuan Wei}, \bibinfo{person}{Jing Sun},
  \bibinfo{person}{Zijiang Zhang}, \bibinfo{person}{Xianhao Zhang},
  \bibinfo{person}{Meng Li}, {and} \bibinfo{person}{Zhe Hou}.}
  \bibinfo{year}{2024}\natexlab{}.
\newblock \bibinfo{title}{{LLM-SmartAudit}: Advanced Smart Contract
  Vulnerability Detection}.
\newblock \bibinfo{howpublished}{CoRR, abs/2410.09381}.
\newblock
\href{https://doi.org/10.48550/arXiv.2410.09381}{doi:\nolinkurl{10.48550/arXiv.2410.09381}}


\bibitem[Wood(2025)]%
        {wood2014ethereum}
\bibfield{author}{\bibinfo{person}{Gavin Wood}.}
  \bibinfo{year}{2025}\natexlab{}.
\newblock \bibinfo{booktitle}{\emph{Ethereum: A Secure Decentralised
  Generalised Transaction Ledger}}.
\newblock \bibinfo{type}{Yellow Paper}. \bibinfo{institution}{Ethereum
  Project}.
\newblock
\urldef\tempurl%
\url{https://ethereum.github.io/yellowpaper/paper.pdf}
\showURL{%
\tempurl}
\newblock
\shownote{Shanghai Version efc5f9a, 2025-02-04; accessed 2026-04-28}.


\bibitem[Wu et~al\mbox{.}(2025)]%
        {wu2025rphunter}
\bibfield{author}{\bibinfo{person}{Hao Wu}, \bibinfo{person}{Haijun Wang},
  \bibinfo{person}{Shangwang Li}, \bibinfo{person}{Yin Wu},
  \bibinfo{person}{Ming Fan}, \bibinfo{person}{Wuxia Jin}, {and}
  \bibinfo{person}{Ting Liu}.} \bibinfo{year}{2025}\natexlab{}.
\newblock \bibinfo{title}{{RPHunter}: Unveiling Rug Pull Schemes in Crypto
  Token via Code-and-Transaction Fusion Analysis}.
\newblock \bibinfo{howpublished}{CoRR, abs/2506.18398}.
\newblock
\href{https://doi.org/10.48550/arXiv.2506.18398}{doi:\nolinkurl{10.48550/arXiv.2506.18398}}


\bibitem[Xia et~al\mbox{.}(2021)]%
        {xia2021trade}
\bibfield{author}{\bibinfo{person}{Pengcheng Xia}, \bibinfo{person}{Haoyu
  Wang}, \bibinfo{person}{Bingyu Gao}, \bibinfo{person}{Weihang Su},
  \bibinfo{person}{Zhou Yu}, \bibinfo{person}{Xiapu Luo}, \bibinfo{person}{Chao
  Zhang}, \bibinfo{person}{Xusheng Xiao}, {and} \bibinfo{person}{Guoai Xu}.}
  \bibinfo{year}{2021}\natexlab{}.
\newblock \showarticletitle{Trade or Trick? Detecting and Characterizing Scam
  Tokens on Uniswap Decentralized Exchange}.
\newblock \bibinfo{journal}{\emph{Proceedings of the ACM on Measurement and
  Analysis of Computing Systems}} \bibinfo{volume}{5}, \bibinfo{number}{3},
  Article \bibinfo{articleno}{39} (\bibinfo{year}{2021}),
  \bibinfo{numpages}{26}~pages.
\newblock
\href{https://doi.org/10.1145/3491051}{doi:\nolinkurl{10.1145/3491051}}


\bibitem[Yuan et~al\mbox{.}(2025)]%
        {yuan2025llmbugscanner}
\bibfield{author}{\bibinfo{person}{Yining Yuan}, \bibinfo{person}{Yifei Wang},
  \bibinfo{person}{Yichang Xu}, \bibinfo{person}{Zachary Yahn},
  \bibinfo{person}{Sihao Hu}, {and} \bibinfo{person}{Ling Liu}.}
  \bibinfo{year}{2025}\natexlab{}.
\newblock \bibinfo{title}{Large Language Model based Smart Contract Auditing
  with {LLMBugScanner}}.
\newblock \bibinfo{howpublished}{CoRR, abs/2512.02069}.
\newblock
\href{https://doi.org/10.48550/arXiv.2512.02069}{doi:\nolinkurl{10.48550/arXiv.2512.02069}}


\bibitem[Zheng et~al\mbox{.}(2023)]%
        {zheng2023judging}
\bibfield{author}{\bibinfo{person}{Lianmin Zheng}, \bibinfo{person}{Wei-Lin
  Chiang}, \bibinfo{person}{Ying Sheng}, \bibinfo{person}{Siyuan Zhuang},
  \bibinfo{person}{Zhanghao Wu}, \bibinfo{person}{Yonghao Zhuang},
  \bibinfo{person}{Zi Lin}, \bibinfo{person}{Zhuohan Li},
  \bibinfo{person}{Dacheng Li}, \bibinfo{person}{Eric~P. Xing},
  \bibinfo{person}{Hao Zhang}, \bibinfo{person}{Joseph~E. Gonzalez}, {and}
  \bibinfo{person}{Ion Stoica}.} \bibinfo{year}{2023}\natexlab{}.
\newblock \showarticletitle{Judging {LLM}-as-a-Judge with {MT}-Bench and
  Chatbot Arena}. In \bibinfo{booktitle}{\emph{Advances in Neural Information
  Processing Systems 36 ({NeurIPS} 2023) Datasets and Benchmarks Track}}.
\newblock
\urldef\tempurl%
\url{https://proceedings.neurips.cc/paper_files/paper/2023/hash/91f18a1287b398d378ef22505bf41832-Abstract-Datasets_and_Benchmarks.html}
\showURL{%
\tempurl}


\bibitem[Zhou et~al\mbox{.}(2023)]%
        {zhou2023sokdefi}
\bibfield{author}{\bibinfo{person}{Liyi Zhou}, \bibinfo{person}{Xihan Xiong},
  \bibinfo{person}{Jens Ernstberger}, \bibinfo{person}{Stefanos Chaliasos},
  \bibinfo{person}{Zhipeng Wang}, \bibinfo{person}{Ye Wang},
  \bibinfo{person}{Kaihua Qin}, \bibinfo{person}{Roger Wattenhofer},
  \bibinfo{person}{Dawn Song}, {and} \bibinfo{person}{Arthur Gervais}.}
  \bibinfo{year}{2023}\natexlab{}.
\newblock \showarticletitle{{SoK}: Decentralized Finance ({DeFi}) Attacks}. In
  \bibinfo{booktitle}{\emph{IEEE Symposium on Security and Privacy (S\&P)}}.
  \bibinfo{publisher}{IEEE}, \bibinfo{address}{San Francisco, CA, USA},
  \bibinfo{pages}{2444--2461}.
\newblock
\href{https://doi.org/10.1109/SP46215.2023.10179435}{doi:\nolinkurl{10.1109/SP46215.2023.10179435}}


\bibitem[Zhou et~al\mbox{.}(2024)]%
        {zhou2024stoppulling}
\bibfield{author}{\bibinfo{person}{Yuanhang Zhou}, \bibinfo{person}{Jingxuan
  Sun}, \bibinfo{person}{Fuchen Ma}, \bibinfo{person}{Yuanliang Chen},
  \bibinfo{person}{Zhen Yan}, {and} \bibinfo{person}{Yu Jiang}.}
  \bibinfo{year}{2024}\natexlab{}.
\newblock \showarticletitle{Stop Pulling my Rug: Exposing Rug Pull Risks in
  Crypto Token to Investors}. In \bibinfo{booktitle}{\emph{Proceedings of the
  46th International Conference on Software Engineering: Software Engineering
  in Practice}}. \bibinfo{publisher}{ACM}, \bibinfo{address}{Lisbon, Portugal},
  \bibinfo{pages}{228--239}.
\newblock
\href{https://doi.org/10.1145/3639477.3639722}{doi:\nolinkurl{10.1145/3639477.3639722}}


\end{thebibliography}

\appendix

\section{Open Science}\label{app:openscience}

This appendix enumerates the artifacts underlying the paper's core
contributions and describes how readers can access them.

\paragraph{Artifacts.}
The following artifacts are provided:
\begin{enumerate}
  \item \textbf{Contract taxonomy:} Source code or documentary
    representations for all 13~contracts; the frozen runtime bytecode used in the evaluation
    (canonical study inputs, with compiler/version metadata preserved); and
    the complete Foundry test suite (71~EVM tests, 13~selector checks).
  \item \textbf{Evaluation pipeline:} Six prompt templates, three judge
    prompts, weighted voting, and the binary verdict mapping logic.
  \item \textbf{Gold-standard labels:} The 50~single-expert cases used for
    judge calibration, their blinded LLM consistency check, plus the broader
    79-case disagreement spot check and 15-case unanimous-agreement sanity
    file that document the labeling audit trail.
  \item \textbf{Complete response database:} All 8{,}528 analyzable model
    responses, plus the 16 refusal responses reported separately, with
    extracted risk ratings where available, per-judge and weighted verdicts,
    and run metadata (model, contract, prompt, repetition).
  \item \textbf{Versioned shepherding extension:} A separate Study-4/5
    delta database, the two held-out contract sources and runtimes, four
    Foundry tests, and scripts that reproduce the held-out and temperature-0
    checks without modifying the frozen 13-contract baseline.
  \item \textbf{Statistical analysis:} Self-contained Python scripts for
    the reported statistics (bootstrap CIs, permutation tests, agreement
    metrics), centered on the artifact reproduction script.
\end{enumerate}

\paragraph{Access.}
All artifacts are permanently archived in Zenodo at:

\begin{center}
  \url{https://doi.org/10.5281/zenodo.22686323}
\end{center}

\noindent The baseline repository has been available since April~29, 2026 and
remains frozen. Shepherding additions are distributed as a separate versioned
extension rather than altering the 13-contract baseline, prompts, or judge
pipeline. This supports longitudinal scoring on identical inputs while allowing
new contracts to be evaluated through the same harness. The artifacts are
intended for research reproduction and defensive evaluation; they are labeled
as synthetic and include local-test instructions rather than deployment guidance.

\paragraph{Artifacts not shared.}
The commercial LLM API keys used during evaluation cannot be shared.
However, all model responses are included in the response database,
enabling full verification of all reported results without re-running
the evaluation. The API access methods (provider, endpoint, model
identifier, reasoning setting, temperature, access window) are
documented in the artifact's model-versioning table, and the code records the exact
provider-side model identifiers and request defaults used by the
harness.

\section{Ethical Considerations}\label{app:ethics}

This study evaluates LLMs on adversarially deceptive smart contracts. We
address potential ethical concerns below.

\paragraph{No deployment of malicious contracts.}
All 13~contracts in this study are custom-built for research purposes and
have never been deployed on any public blockchain (mainnet or testnet). No
real user funds were at risk at any point during the study. All EVM
execution tests were conducted on local Foundry instances with no network
connectivity.

\paragraph{No novel exploit class.}
The deception techniques employed in our vectors---misleading strings,
multi-hop call chains, XOR-masked selectors, storage indirection, embedded
prompt injections---are documented attack patterns observed in real-world
rug pulls and malicious contracts. Our contribution is the systematic
evaluation of LLM robustness against these techniques, not the invention of
new attack primitives. We do not disclose any vulnerability in production
smart contracts or deployed systems. The released artifacts are
research-labeled, local-test-oriented, and do not include deployment
instructions for public chains.

\paragraph{Responsible use of LLM APIs.}
All API-accessed models were used under the providers' standard terms of
service. We did not use jailbreak prompts, adversarial
persuasion, or follow-up prompts intended to bypass safety mechanisms.
The prompt injection vectors (HV-PI, HV-PI2) embed adversarial strings in
the \emph{analysis input} rather than in the system prompt or evaluation
harness; they test the model's analytical robustness against
artifact-embedded instructions, not its safety alignment.

\paragraph{No human subjects.}
This study evaluates LLM outputs, not human behavior. No human subjects
were involved beyond the domain expert who produced gold-standard labels.
The blinded consistency check re-classified model outputs and did not
involve human participants. No IRB approval was required.

\section{Generative AI Usage}\label{app:aiusage}

In accordance with ACM's authorship policy and the CCS~2026 policy on
generative AI, we disclose all uses of LLM tools in this work.

\paragraph{Contract development.}
Claude Opus (Anthropic) was used as a code-generation assistant during
the iterative development of the deception vector contracts
(Section~\ref{sec:taxonomy:vectors}). Each generated contract was
validated through (1)~manual source-code review, (2)~compilation and
freezing of the final runtime bytecode recorded in the artifact,
(3)~the full Foundry EVM test suite (71~tests, all passing), and
(4)~bytecode-level selector verification. The author reviewed and
modified all generated code; no contract entered the evaluation set
without the applicable validation steps.

\paragraph{LLM-as-judge pipeline.}
Three LLMs (DeepSeek Reasoner, Gemini Flash, Ministral-14B) serve as
automated judges in the verdict classification pipeline
(Section~\ref{sec:method:judge}). Their outputs were calibrated against
50~human gold-standard labels with documented per-category accuracy, and
the weighted voting scheme was designed based on measured judge
complementarity. All judge verdicts are included in the released
response database for independent verification.

\paragraph{Blinded consistency check.}
A separate Claude instance, blinded to the primary rater's labels and
justifications, re-coded three rounds of gold-standard cases
(15+15+20 cases) using the same operationalized rubric and matched the
primary labels in 49 of 50~cases (98\% agreement; the per-round breakdown
and single disagreement are documented in the released artifact). We treat this as a reproducible
consistency check on the coding rules, not as a substitute for an
independent human second annotator; that limitation remains acknowledged
in Section~\ref{sec:discussion}.

\paragraph{Frontier-judge re-validation.}
Six frontier LLMs from Anthropic, OpenAI, and Google (full identifiers in
the released artifact) were used as out-of-stack judges to
re-classify the 50-case calibration set under a four-question protocol
distinct from the production-judge prompt. This serves as an independent
robustness check on the headline detection effect; the production stack's
verdicts and the artifact-released response database remain authoritative.

\paragraph{Paper preparation.}
Claude (Anthropic) was used for editorial assistance during paper
drafting, including restructuring of sections, refinement of statistical
language, and grammar checking. OpenAI Codex was used to locate the accepted
source, implement camera-ready formatting, and check compilation and layout.
All substantive claims, interpretations, and analytical conclusions were
produced and verified by the author.

\end{document}